\documentclass[11pt,a4paper]{article}
\pdfoutput=1
\usepackage[utf8]{inputenc}
\usepackage[T1]{fontenc}
\usepackage[british]{babel}
\usepackage[table,svgnames]{xcolor}
\usepackage{%
    lmodern,
    graphicx,
    subcaption,
    enumitem,
    siunitx,
    multirow,
    booktabs,
    xifthen,
    xstring,
    pdflscape,
    longtable
}
\usepackage[floatrowsep=columnsep]{floatrow}
\usepackage{jheppub}
\usepackage{mathabx}
\usepackage{cleveref}

\crefname{section}{section}{sections}
\Crefname{section}{Section}{Sections}
\crefname{figure}{figure}{figures}
\Crefname{figure}{Figure}{Figures}
\crefname{table}{table}{tables}
\Crefname{table}{Table}{Tables}
\crefname{appendix}{appendix}{appendices}
\Crefname{appendix}{Appendix}{Appendices}
\crefname{equation}{eq.}{eqs.}
\Crefname{equation}{Eq.}{Eqs.}
\crefname{enumi}{step}{steps}
\Crefname{enumi}{Step}{Steps}

\newcommand{\onlinecite}[1]{%
    \IfSubStr{#1}{,}{refs}{ref}.~\citealp{#1}%
}
\newcommand{\Onlinecite}[1]{%
    \IfSubStr{#1}{,}{Refs}{Ref}.~\citealp{#1}%
}

\setenumerate{label=(\roman*),itemsep=0pt}
\graphicspath{{figures/}}
\newcommand{\tabEdge}{\hspace{2mm}}
\newsavebox{\largestimage} 

\newcommand{\order}[2][]{\mathcal{O}#1(#2#1)}
\newcommand{\plane}[2]{(#1,\:\!#2)}
\newcommand{\coeff}[2]{\mathcal{\uppercase{#1}}_{#2}}
\newcommand{\chidof}{\chi^2_\text{d.o.f.}}
\newcommand{\crit}{c}
\newcommand{\tric}{\text{tric}}

\newcommand{\clqcd}{\texttt{CL\kern-.25em\textsuperscript{2}QCD}}
\newcommand{\Ocl}{OpenCL}
\newcommand{\Amd}{AMD}
\newcommand{\bahamas}{\texttt{BaHaMAS}}

\newcommand{\NSigma}{N_\sigma}
\newcommand{\NTau}[1][]{%
    \ifthenelse{\isempty{#1}}%
      {N_{\tau}}%
      {N_{\tau_{#1}}}%
}
\newcommand{\Nf}{N_\text{f}}
\newcommand{\T}{T}
\newcommand{\mud}{m_{u,d}}
\newcommand{\ms}{m_{s}}
\newcommand{\Tc}{\T_{\crit}}
\newcommand{\mc}{m_{\crit}}
\newcommand{\betac}{\beta_{\crit}}
\newcommand{\kappac}{\kappa_{\crit}}

\newcommand{\mPS}{m_{PS}}
\newcommand{\mPSc}{\mPS^{\crit}}

\newcommand{\psibarpsi}{\bar{\psi}\psi}
\newcommand{\chiralcond}{\langle \psibarpsi \rangle}
\newcommand{\Action}{\mathcal S}
\newcommand{\Skewness}{B_3}
\newcommand{\Kurtosis}{B_4}

\title{On the order of the QCD chiral phase transition for different numbers of quark flavours }
\author{Francesca Cuteri, Owe Philipsen, Alessandro Sciarra}
\affiliation{Institut f\"ur Theoretische Physik, \\
Goethe-Universit\"at Frankfurt am Main, \\ Max-von-Laue-Str.~1, 60438 Frankfurt am Main, Germany}

\emailAdd{cuteri@itp.uni-frankfurt.de}
\emailAdd{philipsen@itp.uni-frankfurt.de}
\emailAdd{sciarra@itp.uni-frankfurt.de}

\abstract{%
  The nature of the QCD chiral phase transition in the limit of vanishing quark masses has remained elusive for a long time, since it cannot be simulated directly on the lattice and is strongly cutoff-dependent.
  We report on a comprehensive ongoing study using unimproved staggered fermions with $\Nf\in[2,8]$ mass-degenerate flavours on \mbox{$\NTau\in\{4,6,8\}$} lattices, in which we locate the chiral critical surface separating regions with first-order transitions from crossover regions in the bare parameter space of the lattice theory.
  Employing the fact that it terminates in a tricritical line, this surface can be extrapolated to the chiral limit using tricritical scaling with known exponents.
  Knowing the order of the transitions in the lattice parameter space, conclusions for approaching the continuum chiral limit in the proper order can be drawn.
  While a narrow first-order region cannot be ruled out, we find initial evidence consistent with a second-order chiral transition in all massless theories with $\Nf\leq 6$, and possibly up to the onset of the conformal window at $9\lesssim\Nf^*\lesssim 12$.
  A reanalysis of already published $\order{a}$-improved $\Nf=3$ Wilson data on $\NTau\in[4,12]$ is also consistent with tricritical scaling, and the associated change from first to second-order on the way to the continuum chiral limit.
  We discuss a modified Columbia plot and a phase diagram for many-flavour QCD  that reflect these possible features.
}

\arxivnumber{2107.12739}
\newcommand{\fitBetaMNtIVchidof}{0.402}
   \newcommand{\coeffCzeroNtIVWithError}{5.257(7)}
   \newcommand{\coeffConeNtIVWithError}{0.04(9)}
   \newcommand{\coeffCtwoNtIVWithError}{-2.41(28)}
\newcommand{\fitBetaMNtVIchidof}{0.964}
   \newcommand{\coeffCzeroNtVIWithError}{5.257(17)}
   \newcommand{\coeffConeNtVIWithError}{-0.07(19)}
   \newcommand{\coeffCtwoNtVIWithError}{-4.6(6)}

\newcommand{\fitMNfNtIVchidof}{0.685}
   \newcommand{\coeffDcenterNtIVWithError}{1.719(24)}
   \newcommand{\coeffDoneNtIVWithError}{0.052(7)}
   \newcommand{\coeffDtwoNtIVWithError}{-0.035(8)}
\newcommand{\fitMNfNtVIchidof}{1.087}
   \newcommand{\coeffDcenterNtVIWithError}{2.23(8)}
   \newcommand{\coeffDoneNtVIWithError}{0.0065(10)}
   \newcommand{\coeffDtwoNtVIWithError}{-0.0018(4)}

\newcommand{\fitMTfirstQuadraticNfVchidof}{53.541}
   \newcommand{\coeffFToneQuadraticNfVWithError}{-0.236(4)}
   \newcommand{\coeffFTtwoQuadraticNfVWithError}{2.301(24)}
\newcommand{\fitMTfirstCubicNfVchidof}{411.177}
   \newcommand{\coeffFTtwoCubicNfVWithError}{-0.467(20)}
   \newcommand{\coeffFTthreeCubicNfVWithError}{7.44(12)}
\newcommand{\fitMTfirstQuadraticNfVIchidof}{5.101}
   \newcommand{\coeffFToneQuadraticNfVIWithError}{-0.142(7)}
   \newcommand{\coeffFTtwoQuadraticNfVIWithError}{2.29(3)}
\newcommand{\fitMTfirstCubicNfVIchidof}{9.856}
   \newcommand{\coeffFTtwoCubicNfVIWithError}{0.88(3)}
   \newcommand{\coeffFTthreeCubicNfVIWithError}{3.40(16)}
\newcommand{\fitMTfirstQuadraticNfVIIchidof}{16.395}
   \newcommand{\coeffFToneQuadraticNfVIIWithError}{-0.084(8)}
   \newcommand{\coeffFTtwoQuadraticNfVIIWithError}{2.43(4)}
\newcommand{\fitMTfirstCubicNfVIIchidof}{20.860}
   \newcommand{\coeffFTtwoCubicNfVIIWithError}{1.60(4)}
   \newcommand{\coeffFTthreeCubicNfVIIWithError}{1.99(20)}

\newcommand{\fitMpsTNtVIIItoXIIchidof}{0.719}
   \newcommand{\coeffHcenterNtVIIItoXIIWithError}{0.0407(11)}
   \newcommand{\coeffHoneNtVIIItoXIIWithError}{5.93(9)}
\newcommand{\fitMpsTNtVItoXIIchidof}{0.011}
   \newcommand{\coeffHcenterNtVItoXIIWithError}{0.0469(18)}
   \newcommand{\coeffHoneNtVItoXIIWithError}{7.14(29)}
   \newcommand{\coeffHtwoNtVItoXIIWithError}{-9.6(1.8)}
\newcommand{\fitMpsTNtIVtoXIIchidof}{3.147}
   \newcommand{\coeffHcenterNtIVtoXIIWithError}{0.0507(5)}
   \newcommand{\coeffHoneNtIVtoXIIWithError}{7.82(6)}
   \newcommand{\coeffHtwoNtIVtoXIIWithError}{-14.1(3)}
   \newcommand{\coeffHcenterNtIVtoVIIIWithError}{0.0532(11)}
   \newcommand{\coeffHoneNtIVtoVIIIWithError}{8.03(11)}
   \newcommand{\coeffHtwoNtIVtoVIIIWithError}{-15.1(5)}

\begin{document}
\maketitle


\clearpage
\section{Introduction}

The order of the QCD chiral phase transition in the limit of massless quarks  has been an outstanding question for decades.
Many aspects of hadrons and their interactions are remnants of the chiral symmetry of QCD, which nature breaks only weakly by the small $u,d$-quark masses.
One therefore expects the thermal crossover~\cite{Aoki:2006we}, which gradually restores chiral symmetry as a function of increasing temperature, to be similarly related to the phase transition in the chiral limit.
Moreover, the phase transition in the chiral limit  severely constrains the possibilities for the physical phase diagram at finite baryon densities (see, e.g., \onlinecite{Rajagopal:2000wf} and references therein).
A definitive, non-perturbative understanding of this transition is therefore mandatory.

Unfortunately, lattice QCD cannot be simulated in the chiral limit because the fermion determinant is singular for vanishing quark masses.
Any numerical study necessarily proceeds by extrapolating results from a sequence of finite masses to the chiral limit, which inevitably introduces systematic uncertainties.
Moreover, the traditional staggered and Wilson discretisations break chiral symmetry partially or entirely, so that the continuum limit has to be taken before the chiral limit can be approached.
On the other hand, lattice actions with chiral symmetry at finite lattice spacing, such as domain wall or overlap fermions, are by an order of magnitude more expensive to simulate.
For these reasons, the nature of the chiral phase transition could not be conclusively settled so far.

In this work, we attempt to reduce some of these limitations by performing a novel and comprehensive analysis of the nature of the chiral phase transition as a function of the number of degenerate quark flavours $\Nf\in[2,8]$ and lattice spacing, $\NTau\in\{4,6,8\}$.
We use the standard Wilson gauge and staggered fermion actions without improvement terms.
On coarse lattices, this discretisation shows clear first-order transitions in the light mass regime, which weaken with increasing bare quark mass to vanish at some critical value.
The bare parameter space of our lattice action, consisting of the lattice gauge coupling $\beta$, bare quark mass $am$, the number of flavours $\Nf$ and the temporal extent of the lattice $\NTau$, then features a three-dimensional region with first-order transitions, bounded by a critical surface separating it from the crossover region, which we systematically map out and trace along decreasing lattice spacing.
Our analysis differs from others in exploiting the crucial fact that this surface has to terminate in a tricritical line in the chiral limit of the lattice theory, to which one can extrapolate using tricritical scaling with known exponents.
Whenever this applies, it eliminates one important source of systematic uncertainty.

Knowing the order of the phase transition in the bare parameter space of the lattice theory, one can draw conclusions about the continuum phase transition when the continuum and chiral limits are taken in the appropriate order.
Surprisingly, our currently available data prefer the continuum chiral phase transition to be second-order for all $\Nf\in[2,6]$, and suggest that it might stay second-order up to the onset of the conformal window at $9\lesssim\Nf^*\lesssim 12$.
As a check of our analysis in an independent discretisation scheme, we reanalyse already published data for $\Nf=3$ $\order{a}$-improved Wilson fermions.
These are fully consistent with tricritical scaling  when approaching the lattice chiral limit, which implies the same conclusion of a second-order chiral transition for $\Nf=3$.

\section{The chiral phase transition in the continuum and on the lattice}

\subsection{The Columbia plot}\label{sec:col}

\begin{figure}[t]
    \centering
    \subcaptionbox{First-order scenario for $\Nf=2, \mud=0$. \label{fig:columbia-first}}{\includegraphics[width=0.48\textwidth]{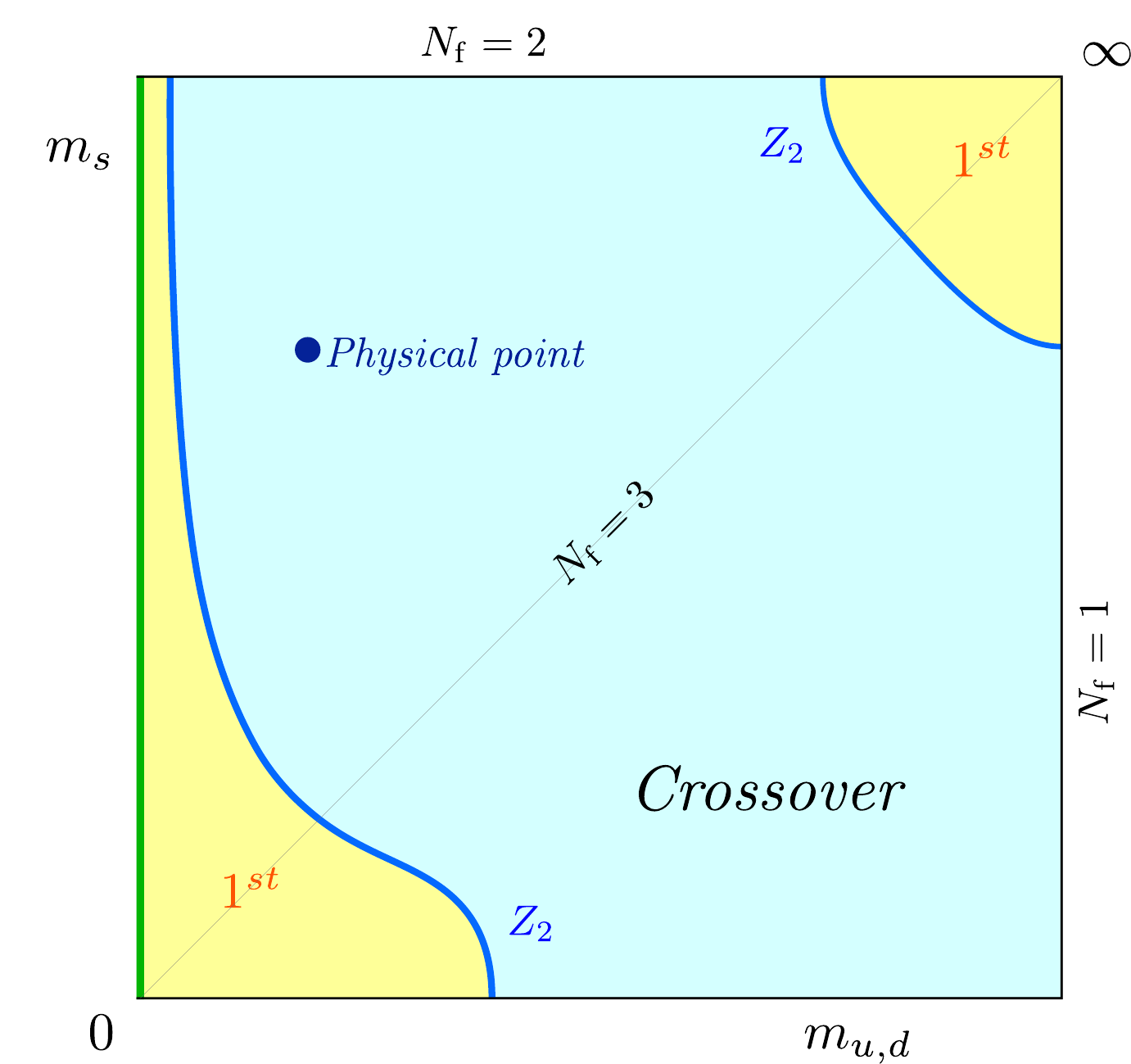}}
    \hfill
    \subcaptionbox{Second-order scenario for $\Nf=2, \mud=0$.\label{fig:columbia-second}}{\includegraphics[width=0.48\textwidth]{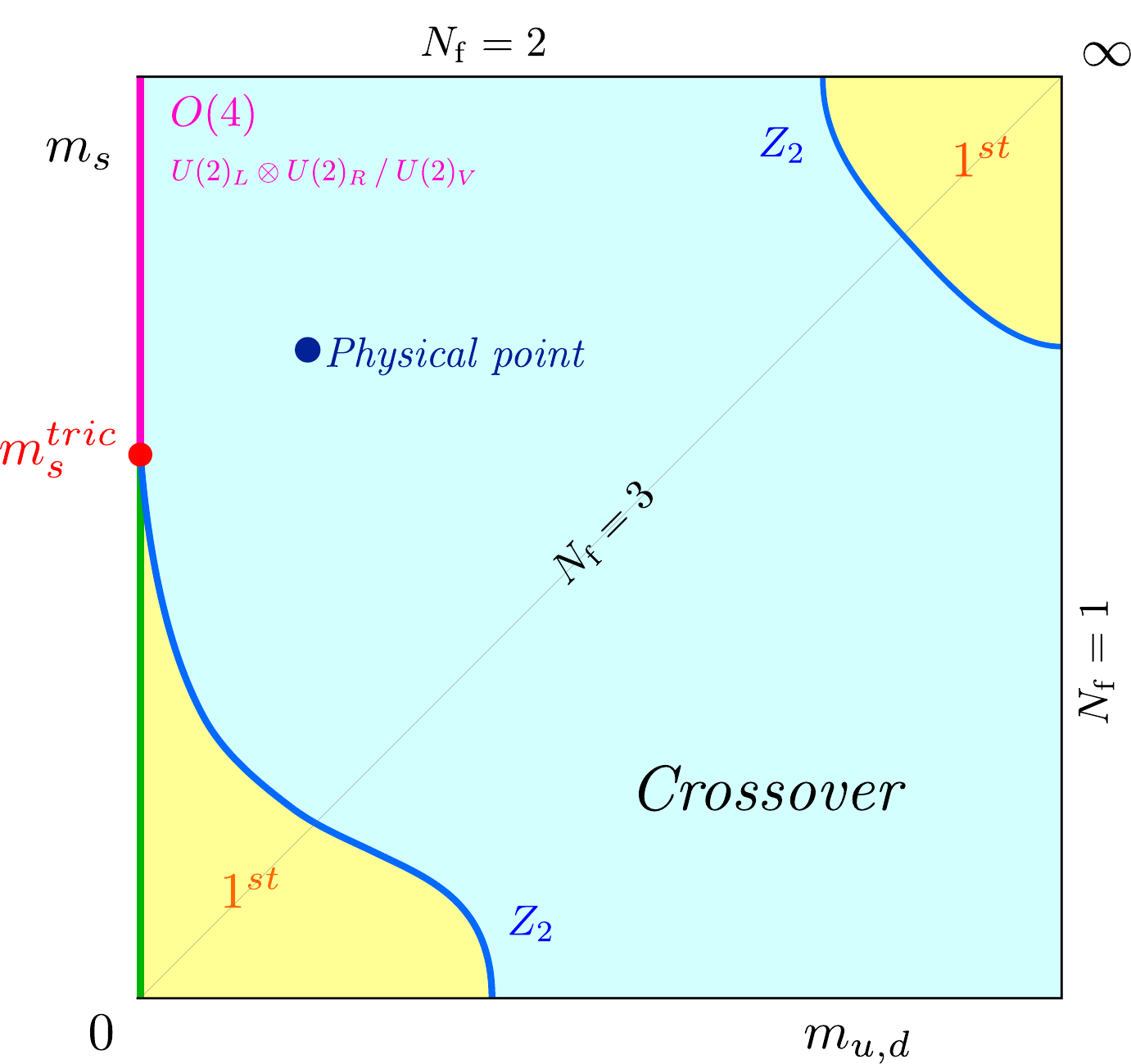}}
    \caption{%
      Possible scenarios for the order of the thermal QCD transition as a function of the quark masses.
      Every point of the plot represents a phase boundary, with an implicitly associated (pseudo-)critical temperature $\Tc(\mud,\ms)$.
    }
    \label{fig:columbia}
\end{figure}

To motivate our analysis by the general picture, we briefly summarise current knowledge about the chiral phase transition.
The thermal QCD transition with physical quark masses has been known for some time to be an analytic crossover~\cite{Aoki:2006we}.
Away from the physical point, the nature of the $\Nf=2+1$  QCD thermal transition as a function of quark masses is usually summarised in a so-called Columbia plot~\cite{Brown:1990ev}, \cref{fig:columbia}.\footnote{Here and in the following,  ``phase boundary'' refers to a (pseudo-)critical combination of parameters irrespective of the nature of the transition, which can be first-order, second-order or crossover.}
In the quenched limit QCD reduces to $SU(3)$ Yang-Mills theory in the presence of static quarks, and shows a first-order phase transition in the continuum limit~\cite{Boyd:1996bx}, associated with the spontaneous breaking of the $Z_3$ centre symmetry.
Finite quark masses break this symmetry explicitly and weaken the first-order phase transition, until it disappears along a critical line in the 3D $Z_2$ universality class.
This region can be simulated directly~\cite{Ejiri:2019csa,Cuteri:2020yke}, and the currently finest lattices predict the critical pseudoscalar mass  for $\Nf=2$ to be about $\mPS\sim 4$ GeV~\cite{Cuteri:2020yke}.

In the chiral limit, the situation is more complicated, because it cannot be simulated.
For a long time expectations have been based on an analysis using the epsilon expansion about $\epsilon=1$  applied to a linear sigma model in three dimensions~\cite{Pisarski:1983ms}.
It predicts the chiral transition to be first-order for $\Nf\geq 3$, whereas the case of $\Nf=2$ is found to crucially depend on the fate of the anomalous $U(1)_A$ symmetry: If the latter remains broken at $\Tc$, the chiral transition should be second order in the $O(4)$-universality class, whereas its effective restoration would enlarge the chiral symmetry and push the transition to first-order.
For non-zero quark masses, chiral symmetry is explicitly broken.
A first-order chiral phase transition then weakens to disappear at a $Z_2$ second-order critical boundary, whereas a second-order transition disappears immediately.
This results in the two scenarios for the Columbia plot depicted in \cref{fig:columbia}.
These early findings were confirmed by numerical simulations of the 3D sigma model~\cite{Gausterer:1988fv} and a high-order perturbative analysis~\cite{Butti:2003nu}.
Yet another analysis of the renormalisation group flow in a 3D effective theory~\cite{Pelissetto:2013hqa} finds a possible symmetry breaking pattern to be $U(2)_L\otimes U(2)_R\rightarrow U(2)_V$ in the case of a restored $U(1)_A$, which would amount to a second-order transition in a different universality class.
Finally, a recent functional renormalisation group analysis applied to $U(1)_A$ restoration in QCD, which is in good agreement with lattice susceptibilities at finite quark masses~\cite{Braun:2020ada}, favours the $O(4)$ scenario in the chiral limit~\cite{Braun:2020mhk}.

A lot of effort has been devoted to disentangle which of these situations is realised in lattice QCD.
There, the bare parameter space of the Columbia plot is enlarged by a third axis for the lattice spacing $a$.
The chiral critical line then traces out a chiral critical surface whose shape is discretisation-dependent, while all valid discretisations must of course merge to the same critical line in the continuum limit.
Early simulations on coarse $\NTau=4$ lattices seemed to confirm the expected first-order transition for $\Nf=3$, both with unimproved staggered~\cite{Karsch:2001nf,deForcrand:2003vyj} as well as standard~\cite{Iwasaki:1996zt} and $\order{a}$-improved~\cite{Jin:2014hea} Wilson fermions.
A narrower first-order region is also seen for $\Nf=2$  with unimproved staggered~\cite{Bonati:2014kpa,Cuteri:2017gci} and unimproved Wilson~\cite{Philipsen:2016hkv} fermions on $\NTau=4$.
However, the location of the boundary line varies widely between these, indicating large cutoff effects.
On the other hand, simulations with an improved staggered action (HISQ) do not see a first-order region on $\NTau=6$ lattices even for $\Nf=3$ down to $\mPS\gtrsim 50$ MeV~\cite{Bazavov:2017xul}.

\begin{figure}[t]
    \centering
    \setlength{\columnsep}{0.04\textwidth}
    \begin{floatrow}
        \ffigbox[0.48\textwidth][][b]{%
            \caption{%
                A Columbia-like plot for the \plane{$a$}{$\mud$}-plane.
                The chiral phase transition for any fixed value of $\ms$ weakens as the lattice spacing is decreased.
            }\label{fig:columbia-like-a-m}
        }{%
            \includegraphics[angle=270, origin=c, width=0.48\textwidth, clip, trim=6mm 0 4mm 0]{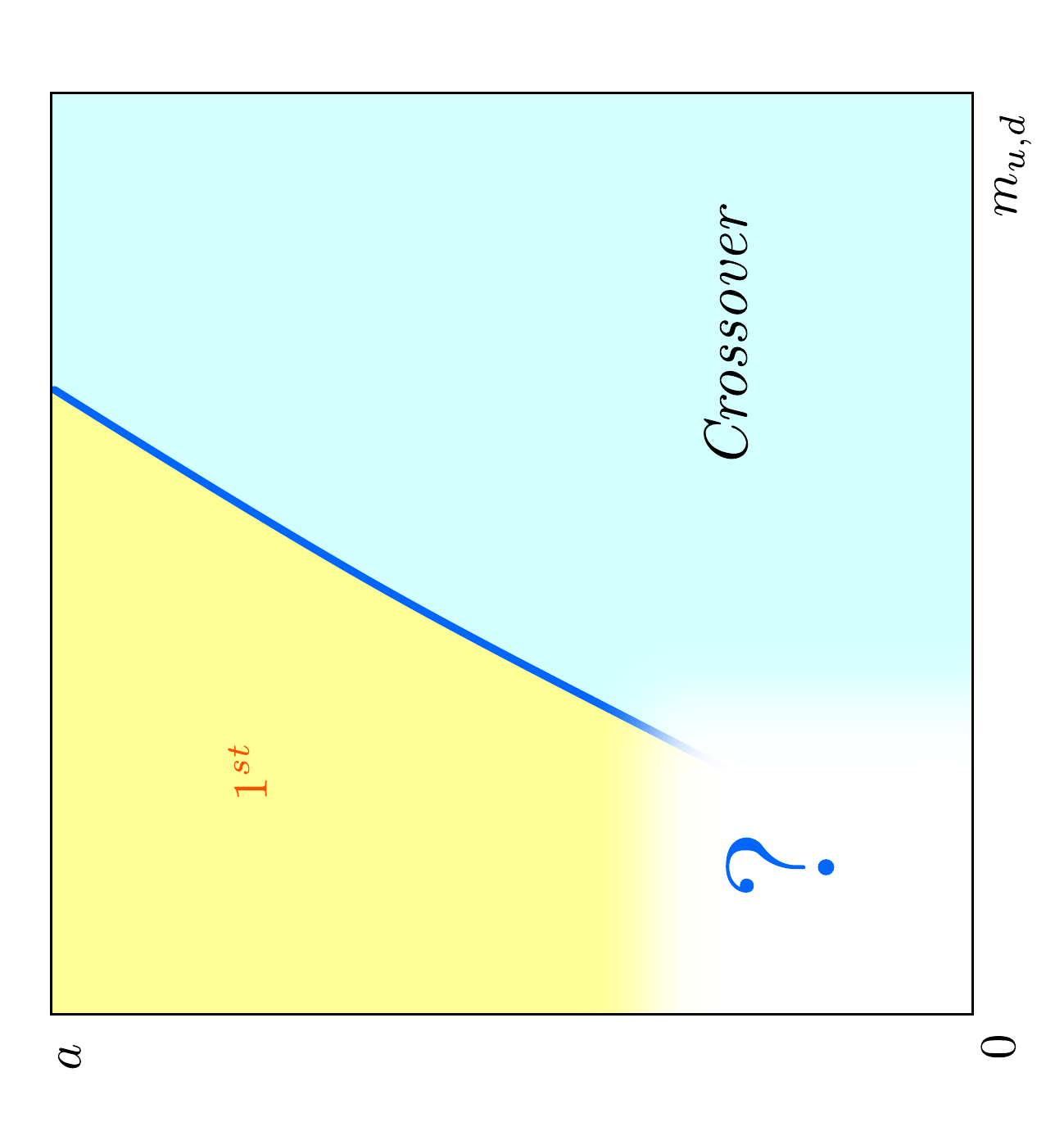}
        }
        \ffigbox[0.48\textwidth][][b]{%
            \caption{%
                Sketch of the $T-\mud$ phase diagram.
                The temperature axis with $T<\Tc$ corresponds to a first-order coexistence line of $\pm\chiralcond\neq 0$, and $\Tc(\mud=0)$ represents a triple point.
            }\label{fig:pbp-3-state}
        }{%
            \includegraphics[width=0.48\textwidth]{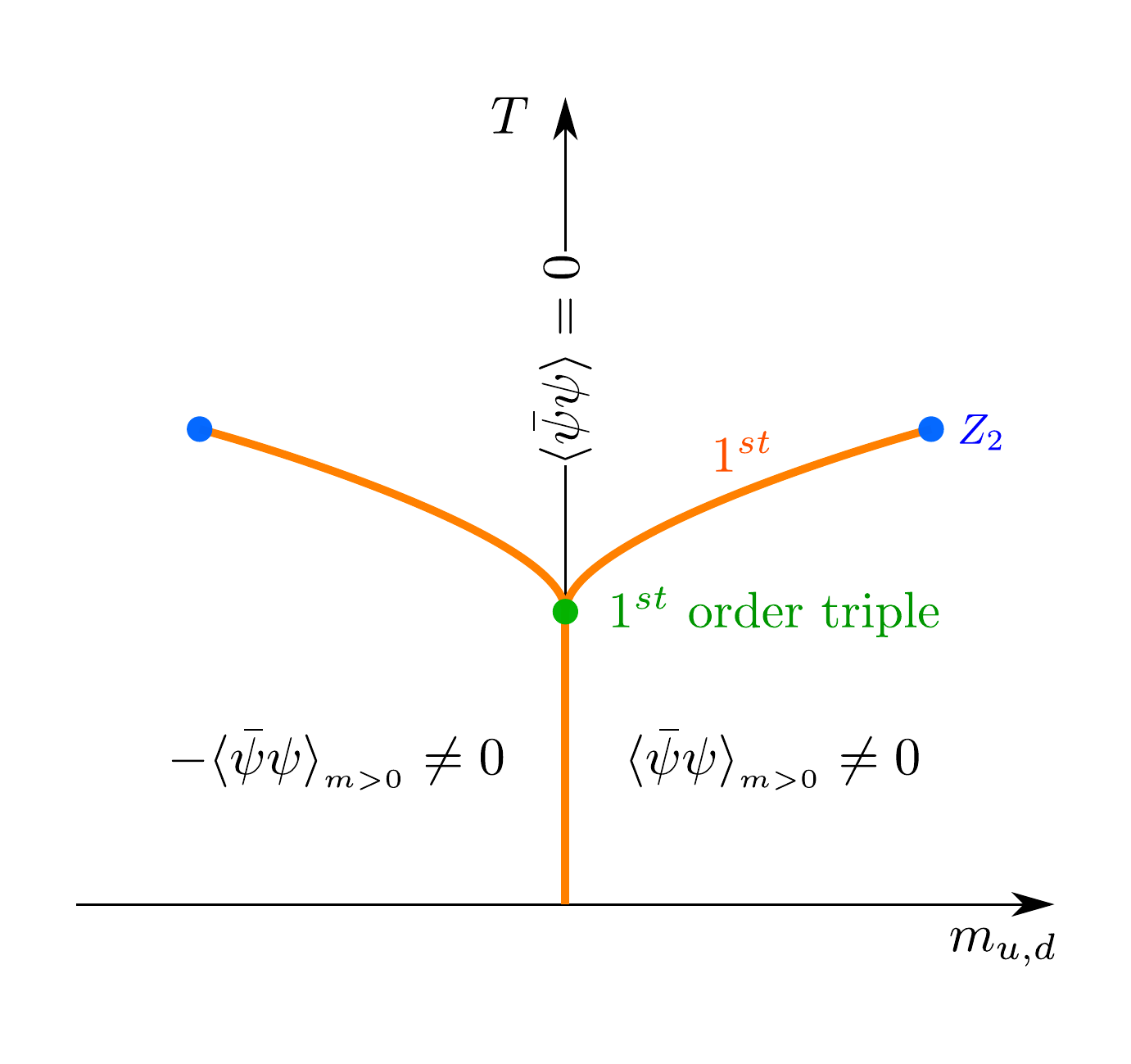}
        }
    \end{floatrow}
\end{figure}

The emerging picture then is as sketched in \cref{fig:columbia-like-a-m} for any fixed strange quark mass, with a shrinking first-order region as the lattice spacing $a$ is decreased.
How much of a first-order region remains in the continuum limit, i.e.~what is the value of the critical quark mass $\mud^c(\ms)$ of the second-order boundary?
It is obvious that any answer to this question strongly depends on the details of extrapolations.
For the direct approach of following the chiral critical line along finer lattices, this has been demonstrated in the case of $\Nf=3$ with $\order{a}$-improved Wilson fermions~\cite{Jin:2014hea,Jin:2017jjp,Kuramashi:2020meg}.
While on the finest $\NTau=12$ lattice the pion mass on the chiral critical point is still $m^c_\pi\approx 110$ MeV, continuum extrapolations give either finite or zero critical masses, depending on their functional form.
Even for $\Nf=4$ the situation is similar, both for unimproved staggered~\cite{deForcrand:2017cgb} and $\order{a}$-improved Wilson quarks~\cite{Ohno:2018gcx}.
For improved staggered fermions, no first-order region is seen at all at the $\Nf$ values simulated so far.
Recently, the critical temperature of the chiral phase transition was determined using HISQ fermions tuned to the physical strange quark mass, simulating a sequence of decreasing light quark masses down to $\sim 58$ MeV in the crossover region~\cite{Ding:2019prx}.
These were extrapolated to the chiral limit employing the well-known scaling relation
\begin{equation}
    \Tc(\mud,\ms)=\Tc(0,\ms)+A(\ms) \;\mud^{1/\beta\delta}\;, \quad \Tc(0,\ms^\text{phys})=132^{+3}_{-6}\;\text{MeV}\;.
\end{equation}
Note that this formula applies in the continuum for infinite volume.
The quoted errors cover the systematic uncertainties obtained by interchanging the order of continuum and chiral extrapolations, as well as using either $O(2)$ or $Z_2$ values for the critical exponents $\beta,\delta$~\cite{Ding:2019prx}.
This statement implies conversely, that an unambiguous distinction between the two scenarios based on this scaling equation is impossible at foreseeable accuracies. A similar investigation with twisted mass Wilson fermions based on slightly larger pion masses confirms this behaviour \cite{Kotov:2021rah}.

\subsection{Three-state coexistence and tricritical scaling}

Some of these obstacles can be circumvented by an alternative analysis, which applies whenever the transition in the massless limit changes between first-order and second-order as a function of some parameter of the theory.
As an example, consider the scenario in \cref{fig:columbia-second}.
A change in the order of the chiral phase transition between $\Nf\in\{2,3\}$ implies the existence of a tricritical point on the strange quark mass axis.
This is because the first-order transitions for $\ms<\ms^\tric$ in the limit $\mud=0$ represent a triple line, with a three-state coexistence (at the critical temperature) characterised by positive, negative and vanishing chiral condensate, cf.~\cref{fig:pbp-3-state}.
As the strange quark mass is increased, the latent heat of the corresponding first-order transition diminishes, until it vanishes in a tricritical point.
This point also marks the confluence of two ordinary critical lines separating crossover and first-order regions for the transitions between $\pm\chiralcond\neq 0$, respectively, and $\chiralcond=0$.
These critical lines enter the tricritical point as a function of the symmetry-breaking scaling field $\mud$ with a known mean field exponent,
\begin{equation}
    \ms^c(\mud)=\ms^\tric + \coeff{a}{1} \cdot \mud^{2/5} + \order[\big]{\mud^{4/5}}\;,
\end{equation}
since the upper critical dimension for a tricritical point is three.
For a general overview of tricritical scaling in and beyond a mean field treatment, see \onlinecite{lawrie}.
The demonstrated existence of a tricritical point would therefore imply the orders of both $\Nf\in\{2,3\}$ chiral phase transitions (but not their universality class).

While the chiral symmetry of staggered fermions is reduced to $O(2)$ compared to $O(4)$ in the continuum, it preserves invariance under rotations $\chiralcond \rightarrow -\chiralcond$ when $a\mud=0$, and hence reproduces a possible triple line and tricritical point also at finite lattice spacing.
In a first attempt with $\Nf=2+1$ on coarse $\NTau=4$ lattices, the chiral critical line was found to be consistent with tricritical scaling~\cite{deForcrand:2006pv}.
Unfortunately, this is inconclusive for the same reasons as described in the last section: A finite portion of the critical line can always be fitted in terms of different polynomial forms, so that a presently impossible accuracy would be required close to the chiral limit in order to get a compelling distinction between the different versions of the Columbia plot reported in \cref{fig:columbia-first,fig:columbia-second}.

\subsection{The chiral phase transition for \texorpdfstring{$\Nf$}{Nf} mass-degenerate flavours}

\begin{figure}[t]
    \centering
    \savebox{\largestimage}{\includegraphics[width=0.48\textwidth]{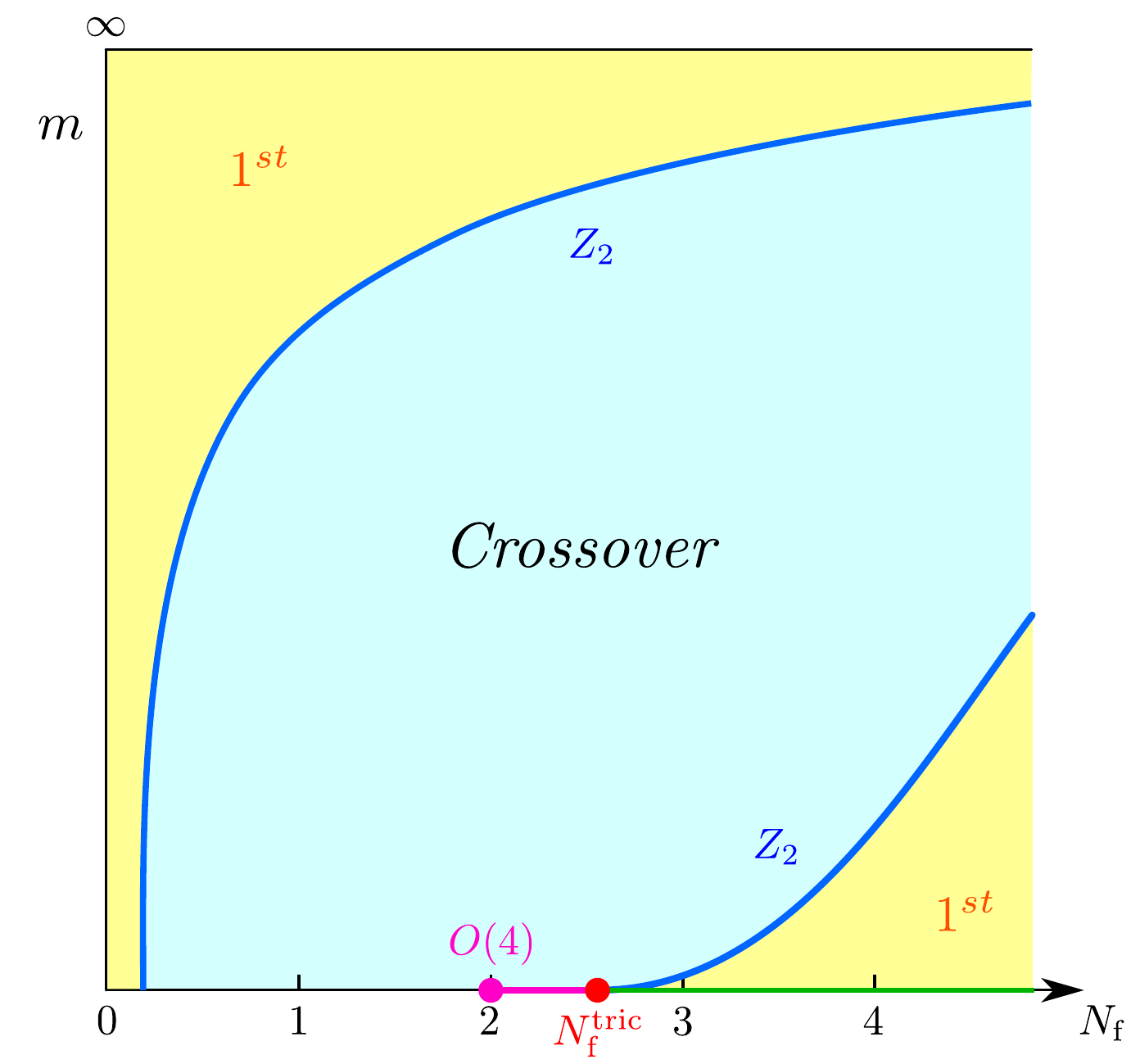}}%
    \setlength{\columnsep}{0.04\textwidth}
    \begin{floatrow}
        \ffigbox[0.48\textwidth][][b]{%
            \caption{%
                Columbia plot for mass-degenerate quarks.
                Every point represents a phase boundary and has an implicitly associated $\Tc(m,\Nf)$.
            }\label{fig:columbia-plot-m-nf}
        }{%
            \usebox{\largestimage}
        }
        \ffigbox[0.48\textwidth][][b]{%
            \caption{%
                Schematic representation of a possible scenario for the $T-m-\Nf$ phase diagram with different numbers of light mass-degenerate flavours.
            }\label{fig:t-m-nf-phase-diagram}
        }{%
            \raisebox{\dimexpr.5\ht\largestimage-.5\height}{%
                \includegraphics[width=0.48\textwidth]{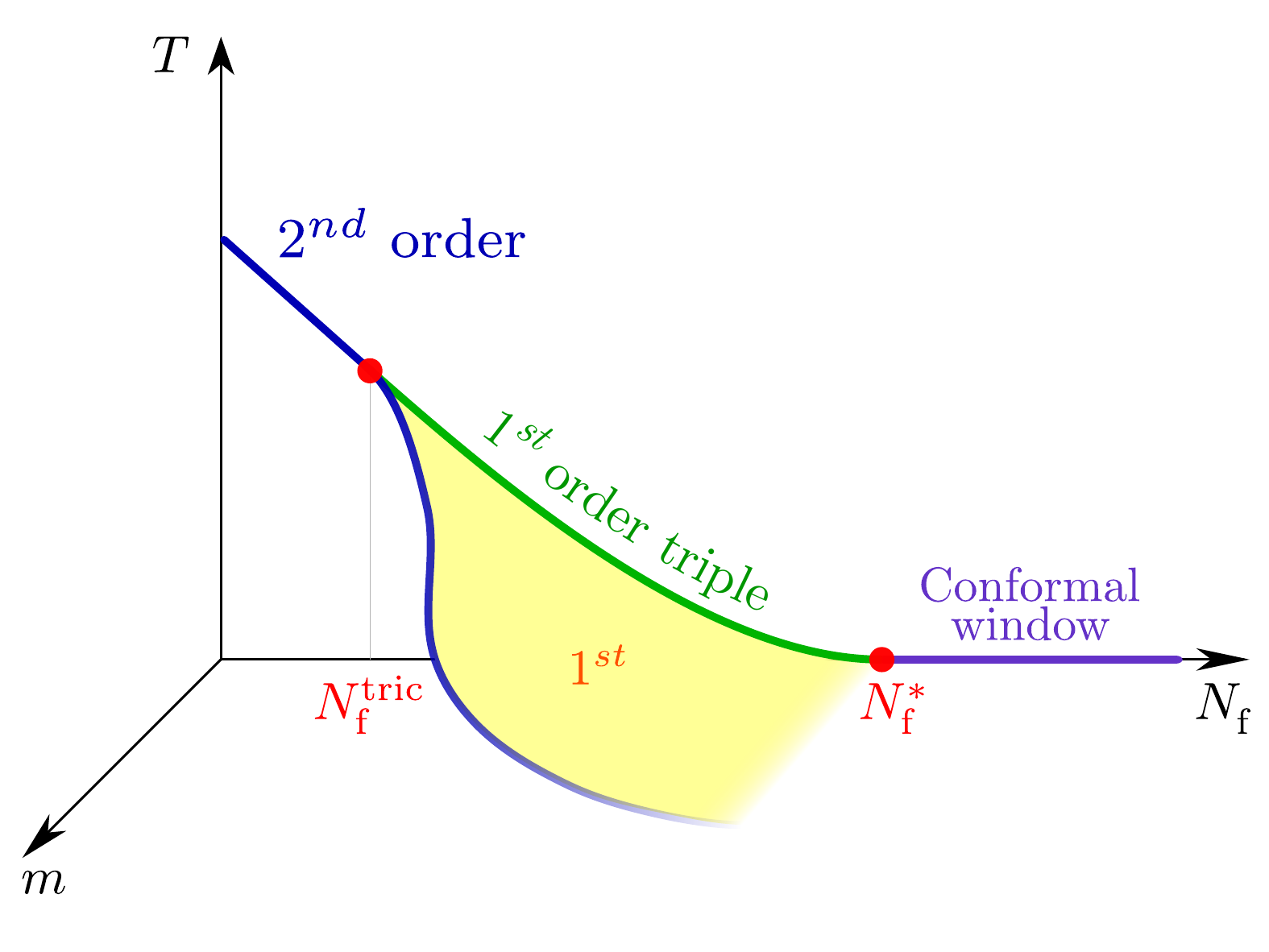}}
        }
    \end{floatrow}
\end{figure}

The way out is to exploit tricritical scaling in a setup, where a tricritical point is guaranteed to exist.
In such a case the scaling form and  its exponents are fixed, and one is only concerned about the location of the tricritical point.
Such a situation emerges from a slight change of perspective and variables, as we suggested previously~\cite{Cuteri:2017gci}.
We now consider degenerate quark masses only, with continuum partition function
\begin{equation}
    Z(\Nf,g,m)=\int {\cal D}A_\mu \; (\det M[A_\mu,m])^{\Nf}\; e^{-\Action_\text{YM}[A_\mu]}.
\end{equation}
Instead of tuning the strange quark mass, an alternative interpolation between $\Nf\in\{2,3\}$, which generalises to larger $\Nf$, is achieved by an analytic continuation of  $\Nf$ to continuous, non-integer values.
In the lattice formulation with rooted staggered fermions, whose determinant is raised to the power $\Nf/4$ in order to describe $\Nf$ mass-degenerate quarks, this is implemented straightforwardly.
The Columbia plot scenario \cref{fig:columbia-second} then translates to the version shown in \cref{fig:columbia-plot-m-nf}, where the tricritical strange quark mass is replaced by a tricritical number of flavours, $2<\Nf^\tric<3$, and the $\Nf$-axis to the right of it corresponds to the new triple line.
The crucial advantage in this modified  parameter space is that, since there is no chiral transition for $\Nf=1$, a tricritical point $\Nf^\tric>1$ is guaranteed to exist as soon as there is a first-order region for any $\Nf>1$.
In particular, the first-order scenario from \cref{fig:columbia-first} now also features a tricritical point, $1<\Nf^\tric<2$.
When a third axis for finite lattice spacing $a$ is added to this picture, there must be a tricritical line $\Nf^\tric(a)$ in the plane $m=0$, which represents the chiral limit of the $Z_2$-critical surface separating lattice parameter regions with first-order transitions from crossover.

The principle of the analysis is now clear: Starting with the already known first-order transitions for $\Nf\in\{3,4\}$ on $\NTau=4$ lattices, map out the $Z_2$ boundary lines until the tricritical scaling region is reached, and extrapolate to the chiral limit,
\begin{equation}\label{eq:scale}
    \Nf^c(am)=\Nf^\tric + \coeff{b}{1} \cdot (am)^{2/5} + \order[\big]{(am)^{4/5}}\;.
\end{equation}
In this way, $\Nf^\tric\approx 1.7$ was obtained on $\NTau=4$ lattices~\cite{Cuteri:2017gci}, implying the first-order scenario for $\Nf=2$.
As a powerful check of the continuation of $\Nf$ as well as tricritical scaling, the critical quark mass for $\Nf=2, \NTau=4$ obtained when keeping $\Nf$ fixed and varying (imaginary) chemical potential~\cite{Bonati:2014kpa} is consistent, based on tricritical scaling, with results at $\Nf\in[2.1,2.2]$~\cite{Cuteri:2017gci}.
In that case the quark mass is again the symmetry breaking scaling field, but with $\Nf\rightarrow (\mu/T)^2$ in \cref{eq:scale}.
In the present work, we systematically extend our study from \onlinecite{Cuteri:2017gci} to larger numbers of flavours and finer lattices.

The chiral phase transition of QCD with many flavours is also of interest in the context of physics beyond the Standard Model.
There the main focus is on the ``conformal window'' $\Nf^*<\Nf<\Nf^\text{a.f.}$, which denotes a range of theories with an infrared fixed point at a finite value of the gauge coupling.
For $\Nf> \Nf^\text{a.f.}$ asymptotic freedom is lost, while $\Nf^*$ is the critical number of flavours marking the endpoint of the existence of a chiral condensate, and hence of chiral phase transitions, as indicated schematically in the possible phase diagram \cref{fig:t-m-nf-phase-diagram}.
A precise non-perturbative determination of this point is difficult for various technical reasons, but is expected in the range $9\lesssim \Nf^* \lesssim 12$.
For an overview of the related strong interaction dynamics and its lattice investigations, see \onlinecite{DeGrand:2015zxa,Nogradi:2016qek,Svetitsky:2017xqk}.

For our present purpose, we stay at $\Nf<\Nf^*$.
Note that the decrease of the critical temperature in \cref{fig:t-m-nf-phase-diagram} is predicted by functional renormalisation group methods~\cite{Braun:2006jd,Braun:2009ns} and lattice simulations~\cite{Miura:2012zqa}, which in the chiral limit implies $\Nf^*$ to represent a quantum phase transition at $T=0$.
An interesting feature of the critical temperature $\Tc(\Nf)$ is its approximately linear behaviour observed in \onlinecite{Braun:2006jd,Braun:2009ns}.
It can be explained by expressing dimensionful quantities in terms of $\Lambda_\text{QCD}(\Nf)$, whose perturbative expression can be expanded in $\Nf$,
\begin{equation}
    \Tc(\Nf)=\Bigg(\frac{\Tc}{\Lambda_\text{QCD}}\Bigg)(\Nf) \cdot \Big[ \Lambda_\text{QCD}(0)+
    \Lambda^{(1)}\,\varepsilon\,\Nf+\order[\big]{(\varepsilon\Nf)^2}\Big]\;.
\end{equation}
The perturbative second factor is indeed approximately linear due to the smallness of $\varepsilon\approx 0.11$, and appears to dominate the remaining $\Nf$-dependence of the dimensionless first factor until the conformal window is approached.
There, non-perturbative scaling associated with the quantum critical point takes over and leads to a bending of the critical line~\cite{Braun:2006jd,Braun:2009ns}.
By contrast, the order of the chiral transition as a function of $\Nf$ is not yet known.
A change from second-order to first-order with increasing $\Nf$ implies the already mentioned tricritical point $\Nf^\tric$.
One would then also expect $\Nf^*$ to be tricritical, as it represents yet another endpoint of a triple line.

\section{Lattice simulations and analysis}

For our numerical investigation, we work with the standard unimproved Wilson gauge and staggered fermion actions.
All numerical simulations have been performed using the publicly available \Ocl-based code \clqcd, which is optimised to run efficiently on \Amd{} GPUs and contains an implementation of the RHMC algorithm for unimproved rooted staggered fermions.
In particular, version \texttt{v1.0}~\cite{pinke_cl2qcd_2018} has been employed for simulations on smaller $\NTau$ on the L-CSC supercomputer, while version \texttt{v1.1}~\cite{sciarra_cl2qcd_2021} has been run on the newer Goethe~HLR supercomputer to run the most costly simulations.
To effectively handle the thousands of necessary simulations, the \bahamas\ software~\cite{sciarra_bahamas_2021} has been extensively used.

Regarding variable numbers of flavours, there is one aspect that is worth mentioning.
It is well known that in the RHMC algorithm, when the condition number of the Dirac matrix is dominated by a few isolated tiny eigenvalues (e.g.\ in the chiral limit), the multiple pseudofermions technique~\cite{Clark2007} is advantageous, since it reduces the extra noise introduced into the system by the stochastic estimate of the fermion determinant.
Using multiple pseudofermions, the maximum fermion force, which might trigger the molecular dynamics integrator instabilities (requiring in turn a smaller integration step size and hence making algorithm more costly), is reduced.
However, when increasing the fermion mass at fixed $\Nf$, the condition number of the Dirac matrix decreases and using more than one pseudofermion turns out to be disadvantageous.
On the other hand, in the present study, we increase the simulated quark mass range and the number of flavours simultaneously.
Since a larger $\Nf$-value also implies larger forces in the RHMC algorithm, it is worth wondering whether using multiple pseudofermions can still help for simulations at larger mass values.
Therefore, the condition number of the Dirac matrix has been monitored in all simulations and it actually turned out that the increase of the fermion force produced by larger $\Nf$ values prevails by far over its decrease due to larger fermion mass values.
Indeed, it sometimes pays off to even increase the number of multiple pseudofermions when simulating at larger $\Nf$, despite $am$ being larger.
\begin{table}[tb]
    \setlength{\tabcolsep}{3mm}
    \renewcommand{\arraystretch}{1.1}
    \centering
    \begin{tabular}{
        @{\tabEdge}
        cl
        S[table-format=1.4]
        S[table-format=1.4]
        @{\hspace{6mm}}
        S[table-format=1.4(2)]
        S[table-format=2]
        S[table-format=1.2]
        S[table-format=3.1]
        @{\hspace{6mm}}
        S[table-format=1.4(2)]
        @{\tabEdge}
        }
        \toprule
        $\NTau$ & $\Nf$ & {$am_\text{min}$} & {$am_\text{max}$}& {$a\mc$} &  {d.o.f.} & {$\chidof$} & {$Q[\%]$} & {$\betac\text{ at }a\mc$}\\
        \midrule
        \multirow{11}{*}{4}
        & 2.1  & 0.0015 & 0.0045 &  0.00343(14)  &   9  &  0.173  &  99.7  &  5.2363(3)  \\
        & 2.2  & 0.0025 & 0.01   &  0.00579(15)  &  10  &  0.257  &  99    &  5.2238(3)  \\
        & 2.4  & 0.0075 & 0.015  &  0.01088(19)  &  13  &  0.603  &  85    &  5.2006(4)  \\
        & 2.6  & 0.0125 & 0.02   &  0.01577(23)  &  10  &  0.230  &  99    &  5.1779(5)  \\
        & 2.8  & 0.0175 & 0.025  &  0.02106(25)  &  10  &  0.270  &  99    &  5.1568(5)  \\
        & 3    & 0.0225 & 0.3    &  0.0264(5)    &  10  &  0.164  &  99.8  &  5.1368(9)  \\
        & 4    & 0.05   & 0.065  &  0.0551(7)    &  10  &  0.365  &  96    &  5.0529(13) \\
        & 5    & 0.07   & 0.09   &  0.0820(8)    &  12  &  0.734  &  72    &  4.9828(15) \\
        & 6    & 0.1    & 0.12   &  0.1078(6)    &   7  &  1.148  &  33    &  4.9234(12) \\
        & 7    & 0.12   & 0.14   &  0.1308(8)    &   7  &  0.874  &  53    &  4.8692(18) \\
        & 8    & 0.14   & 0.17   &  0.1539(11)   &   7  &  0.668  &   7    &  4.8233(24) \\
        \midrule
        \multirow{7}{*}{6}
        & 3    & 0.0025 & 0.005  &  0.0025(3)    &  2  &  1.298 &  27    & 5.2126(12) \\
        & 3.6  & 0.0075 & 0.0125 &  0.00910(20)  &  7  &  1.930 &   6    & 5.1396(7)  \\
        & 4    & 0.0125 & 0.015  &  0.01392(16)  &  7  &  1.482 &  17    & 5.0953(5)  \\
        & 4.4  & 0.015  & 0.025  &  0.0184(4)    &  6  &  2.029 &   6    & 5.0516(12) \\
        & 5    & 0.02   & 0.03   &  0.02614(28)  &  6  &  0.884 &  51    & 4.9931(9)  \\
        & 6    & 0.035  & 0.045  &  0.04012(26)  &  6  &  2.478 &   2    & 4.9081(9)  \\
        & 7    & 0.05   & 0.06   &  0.05372(27)  &  7  &  3.616 &   0.1  & 4.8309(9)  \\
        \midrule
        \multirow{3}{*}{8}
        & 5    & 0.005  & 0.0075 &  0.00608(13)  &  3    &  2.181  &  9     & 4.9828(15) \\
        & 6    & 0.01   & 0.015  &  0.0125(25)   &  {--} &  {--}   &  {--}  &  4.847(15)\\
        & 7    & 0.015  & 0.02   &  0.0175(25)   &  {--} &  {--}   &  {--}  &  4.731(14)\\
        \bottomrule
    \end{tabular}
    \caption{%
      Overview of the finite size scaling analysis and of the $\betac$ interpolation.
      Large values of $\chidof$ are due to the low number of simulated mass values and/or to the still low accumulated statistics.
      For $\Nf\in\{6,7\}$ at $\NTau=8$, the available data are not yet sufficient to obtain an acceptable fit, and the quoted critical mass represents the middle point of simulated masses at which the order of the transition was clear from the ordering of the kurtosis on increasing spatial lattice size, i.e.~towards the thermodynamic limit.
    }
    \label{tab:fits_overview}
\end{table}

Our goal is to determine the location and order of chiral phase transitions in the four-dimensional space spanned by the dimensionless parameters of our lattice action: The lattice gauge coupling $\beta$, the bare quark mass in lattice units $am$, the number of degenerate quark flavours $\Nf$, and the number of time-slices $\NTau$.
For any fixed value of $\NTau$ and $\Nf$, we achieve this by making use of two particular standardised moments,
\begin{equation}
    B_n(\beta,am,\NSigma) =
    \frac{\left\langle\left(\mathcal{O} - \left\langle\mathcal{O}\right\rangle\right)^n\right\rangle}{\left\langle\left(\mathcal{O} - \left\langle\mathcal{O}\right\rangle\right)^2\right\rangle^{{n/2}^{\vphantom{x}}}} \;,
\end{equation}
where the chiral condensate  has been chosen as observable, $\mathcal{O}=\psibarpsi$, as it becomes the order parameter of the thermal phase transition in the chiral limit.
In particular, to extract the order of the transition as a function of the quark mass, we evaluate the kurtosis $\Kurtosis(\betac,am,\NSigma)$~\cite{Binder:1981sa} of the sampled $\chiralcond$ distribution, where $\betac$ denotes the (pseudo-) critical coupling of the phase boundary, for which the zero-skewness condition \mbox{$\Skewness(\beta=\betac,am,\NSigma)=0$} holds.
In the thermodynamic limit $\NSigma \rightarrow \infty$, the kurtosis $\Kurtosis(\betac,am,\NSigma)$ takes the values of 1 for a first order transition and 3 for an analytic crossover, respectively, with a discontinuity when passing from a first order region to a crossover region via a second order point; for the 3D Ising universality class of interest here, it takes the value $1.604$~\cite{Pelissetto:2000ek}.
On finite, increasing  volumes this discontinuity is smoothed out and approached gradually with a rate characteristic of the universality class in question,
\begin{equation}
    \Kurtosis(\betac,am,\NSigma) \approx 1.604 + c\,(am - a\mc)\,\NSigma^{1/0.6301}\quad\text{with}\quad c\in\mathbb{R}\;.
\end{equation}
Data have been produced and analysed in a completely analogous way to that explained in \onlinecite{Cuteri:2017gci,Cuteri:2018wci} and, in particular, the critical mass $a\mc$ has been extracted at fixed $\NTau$ and $\Nf$ by fitting the kurtosis data according to this finite size scaling formula.
However, some steps in the analysis procedure have been improved and will be discussed next.
This is also the reason why older results are reported here slightly changed.

First of all the evaluation of centred moments of observables that are stochastically estimated might be biased, if carried out simply using all stochastic sources for all powers of the observable.
A bias reduction technique consisting of not mixing estimates coming from the same stochastic source when calculating powers of $\psibarpsi$ can be applied in order to figure out how much the naive estimate is biased.
This has been implemented for an arbitrary number of stochastic estimates and has been used in the analysis of all data.\footnote{Thanks to Christian Schmidt for pointing out how to implement the bias reduction technique in a completely general way.}
Interestingly enough, significant corrections to $\Skewness$ and $\Kurtosis$ have been found on single chains, but these are smoothed out when it comes to analyse the merged chains.

Since the subsequent analysis presented in \cref{sec:critical_surface} heavily relies on the outcome of the $\Kurtosis$-fits, we decided to improve the error estimate on $a\mc$ using a more accurate procedure.
Values of $\Kurtosis(\betac,am,\NSigma)$ are obtained using the multiple-histogram method~\cite{FSReweighting}, and their error  is in turn calculated out of hundreds of bootstrap estimators.
Each of these can be used as new central value of $\Kurtosis$, so that a new $\Kurtosis$-data set can be built and fitted to extract a new (bootstrap estimator of) $a\mc$.\footnote{The error on each new $\Kurtosis$ central value is kept fixed to that obtained from the multiple-histogram method.}
Combining all these estimators of $a\mc$, a bootstrap error can be calculated and compared to that coming from the $\chi^2$-minimisation.
This method has been run on all data sets and it often leads to smaller errors on the critical $Z_2$ mass.
The outcome of all fits can be found in \cref{tab:fits_overview}, where also the simulated mass range has been included.
The uncertainties on $a\mc$ are those obtained with our new bootstrap-fitting technique.
In \cref{sec:appendix} a detailed overview of the simulations can be found.
To give an idea of the numerical effort: Over $600$ values of $\beta$ have been simulated in total, producing about $120$ millions of trajectories.

Finally, we explain how the value of $\betac$ at $a\mc$ reported in the last column of \cref{tab:fits_overview} has been extracted.
Rather than running further costly simulations at the extracted critical masses, the values $\betac$ found on the largest simulated $\NSigma$ for each mass value have been linearly interpolated in order to extract $\betac$ at $a\mc$.
In order to attribute an uncertainty to this interpolated value, the data have been also interpolated at $a\mc\pm \sigma_{a\mc}$, namely at the critical mass plus or minus one standard deviation.

\section{The chiral critical surface in the parameter space of staggered fermions}\label{sec:critical_surface}

For the interpretation of the lattice data, it is useful to recall the structure of the four-dimensional parameter space, $\{\beta,am,\Nf,\NTau\}$ in the limit of infinite spatial volume.
The discussion is facilitated by assuming some analytic continuation of both $\Nf$ and $\NTau$ to real values, so that all parameters can be regarded as continuous and treated on equal footing.
The phase boundary associated with chiral symmetry restoration, which we define by vanishing skewness, $\Skewness=0$, then represents a three-dimensional parameter region.
That subspace consists of a region with first-order transitions and a crossover region, separated by a chiral critical surface, on which the kurtosis assumes its 3D Ising value, $\Kurtosis=1.604$.
The identification of this  chiral critical surface in the bare parameter space of the standard staggered lattice action is the central result of our simulations.
In the following we analyse this surface projected onto different variable pairings.

\subsection{Plane of gauge coupling and quark mass}

\begin{figure}[t]
    \centering
    \subcaptionbox{Projections with fixed $\NTau$.\label{fig:beta_m-fix-nt}}{\includegraphics[width=0.495\textwidth]{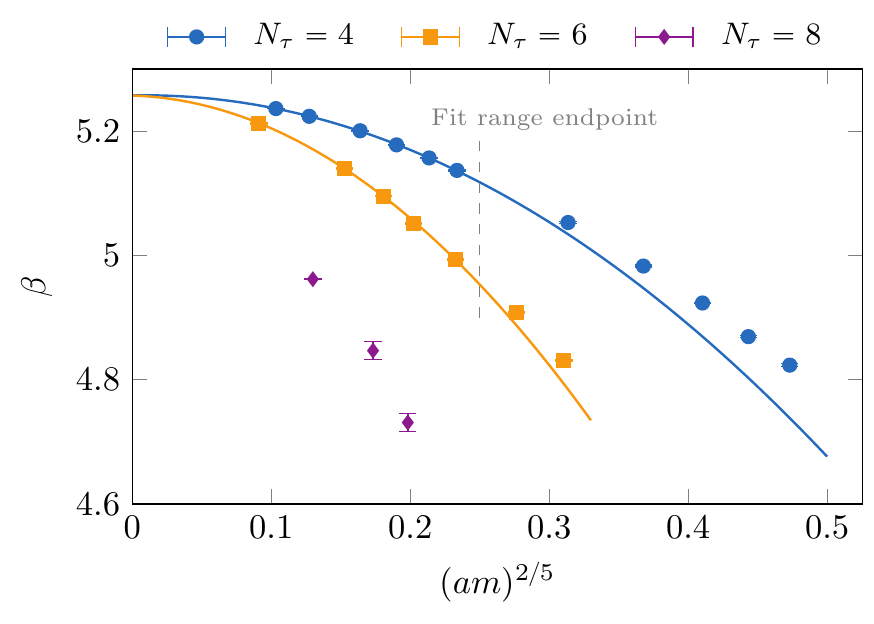}}
    \hfill
    \subcaptionbox{Projections with fixed $\Nf$.\label{fig:beta_m-fix-nf}}{\includegraphics[width=0.495\textwidth]{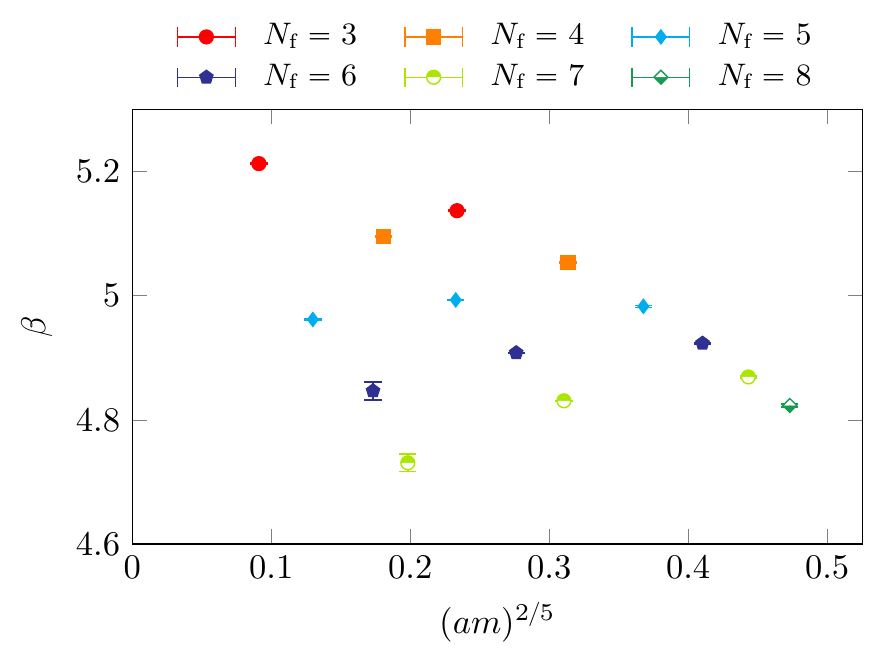}}
    \caption{%
      The chiral critical surface, separating 3D parameter regions with a first-order transition from crossover regions, projected onto the $\plane{\beta}{am}$-plane.
      Every point represents a phase boundary with an implicitly tuned $\betac(am,\Nf,\NTau)$.
      In \subref{fig:beta_m-fix-nt} the lines represent next-to-leading order fits in the scaling variable, cf.~\cref{tab:beta_fit}.
    }
    \label{fig:beta_m}
\end{figure}

The chiral critical surface projected onto the $\plane{\beta}{am}$-plane is shown in \cref{fig:beta_m}.
As in the case of the Columbia plot, every point of \cref{fig:beta_m} represents a point of the phase boundary, with the fourth parameter implicitly determined by the other three.
In \cref{fig:beta_m-fix-nt}, we show slices of the phase boundary corresponding to different values of $\NTau$.
The curve for every given $\NTau$ represents the critical bare parameter combinations with second-order phase transitions, and is parametrised by different values of $\Nf$, increasing from left to right.
The region below the curve corresponds to first-order transitions, whereas the region above the curves correspond to analytic crossover behaviour.
Obviously, this qualitative picture is unchanged when it is restricted to a grid with the physically meaningful integer values for $\Nf$, as in \cref{fig:beta_m-fix-nf}.
However, in that case the underlying functional behaviour is less constrained and difficult to extract.
By filling in non-integer values of $\Nf$, a convincing picture of tricritical scaling is obtained: For each given $\NTau$, the lower $\beta$-axis represents a triple-line of first-order transitions in the chiral limit of the lattice action, which weakens as $\beta$ is increased.
This first-order chiral transition disappears in a tricritical point, from which the line of ordinary $Z(2)$ critical points, $\betac(am)$, emerges.
The critical curve is excellently described over a wide mass range by a next-to-leading order fit in the scaling variable, cf.~\cref{tab:beta_fit},
\begin{equation}\label{eq:beta_fit}
    \betac(am,\Nf(\NTau),\NTau)
    = \beta_\tric(\NTau) 
    + \coeff{c}{1}(\NTau)(am)^{2/5}
    + \coeff{c}{2}(\NTau)(am)^{4/5} + \order[\big]{(am)^{6/5}} \; .
\end{equation}

\begin{table}[t]
    \setlength{\tabcolsep}{3mm}
    \renewcommand{\arraystretch}{1.1}
    \centering
    \begin{tabular}{
        @{\tabEdge}
        cc
        S[table-format=1.3(2)]S[table-format=+1.2(2)]S[table-format=+1.2(2)]
        @{\hspace{8mm}}
        S[table-format=1.2, round-mode = places, round-precision = 2]
        @{\tabEdge}
      }
      \toprule
      $\NTau$ & range in $(am)^{2/5}$ & {$\beta_\tric$} & {$\coeff{c}{1}$} & {$\coeff{c}{2}$} & {$\chidof$} \\
      \midrule
      4 & $[0,0.25]$ &  \coeffCzeroNtIVWithError  &  \coeffConeNtIVWithError  & \coeffCtwoNtIVWithError & \fitBetaMNtIVchidof \\
      6 & $[0,0.25]$ &  \coeffCzeroNtVIWithError  &  \coeffConeNtVIWithError  & \coeffCtwoNtVIWithError & \fitBetaMNtVIchidof \\
      \bottomrule
    \end{tabular}
    \caption{Fits of the data in \cref{fig:beta_m-fix-nt} to \cref{eq:beta_fit}.}
    \label{tab:beta_fit}
\end{table}

\Cref{fig:beta_m-fix-nf} shows the same chiral critical surface, this time sliced at fixed $\Nf$, so that the curves are implicitly parametrised by $\NTau$, increasing from right to left.
Again, for each $\Nf$ the parameter region below the curves describes first-order transitions and the region above it analytic crossovers.
However, with presently only three $\NTau\in\{4,6,8\}$, a functional behaviour is difficult to assess.
This further illustrates the advantage that the additional non-integer $\Nf$ values offer in \cref{fig:beta_m-fix-nt}.
Note that a finer resolution in $\NTau$ could in principle be achieved by working with anisotropic lattices, but this would come at the price of introducing a second lattice gauge coupling, and thus another dimension to the parameter space of the lattice theory.

Clearly, \cref{fig:beta_m-fix-nt,fig:beta_m-fix-nf} show the same chiral critical surface, presented by slicing the three-dimensional subspace of the phase boundary in different directions.
In particular, the bare quark mass $am$ is the symmetry breaking field and, sufficiently close to the chiral limit, behaves as a tricritical scaling variable in any plane that contains a tricritical point and is not orthogonal to the $am$-axis.
This is guaranteed to be the case for every curve in \cref{fig:beta_m-fix-nt}, where all $\Nf$ are included in every fixed $\NTau$ curve, but not in \cref{fig:beta_m-fix-nf}, since for a given fixed $\Nf$ there need not be a tricritical point.
The chiral extrapolations shown in \cref{fig:beta_m-fix-nt} then lead to a one-dimensional tricritical line,
\begin{equation}
    \beta_\tric(\NTau) = \betac(0,\Nf^\tric(\NTau), \NTau)\;.
\end{equation}
With a non-perturbative scale setting for the lattice spacing $a(\beta)$, $\beta_\tric$ could be converted into a tricritical temperature $T_\tric(\NTau)=T_\tric(a)$, corresponding to the tricritical point in \cref{fig:t-m-nf-phase-diagram}, displaced due to the finite lattice spacing.
This is not easily feasible in practice, because a scale would have to be set for arbitrary values of ${\Nf}$ in the chiral limit.

Fortunately, we do not need this temperature for our present purpose.
Here, we conclude that there is unambiguous evidence for the chiral critical surface terminating in a tricritical line in the chiral limit of the lattice theory, which it approaches as a function of the scaling field $(am)^{2/5}$.
With this knowledge, we can now analyse the chiral critical surface in different pairings of variables.

\subsection{Plane of quark mass and number of flavours}

\begin{figure}[t]
    \centering
    \subcaptionbox{Projection onto the $\plane{am}{\Nf}$-plane.\label{fig:m_nf-fixed-nt}}{\includegraphics[width=0.495\textwidth]{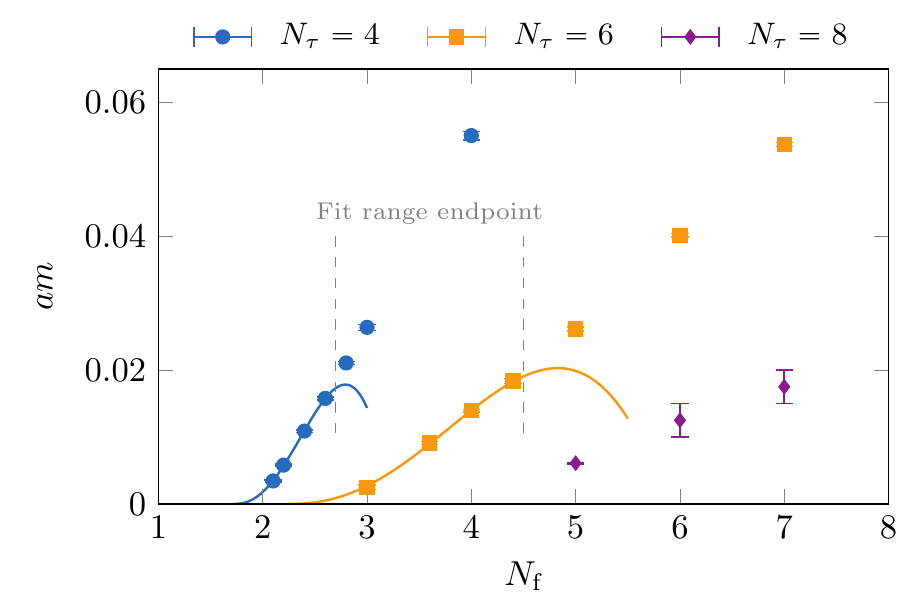}}
    \hfill
    \subcaptionbox{Projection onto the $\plane{am}{\NTau^{-1}}$-plane.\label{fig:m_nt-fixed-nf}}{\includegraphics[width=0.495\textwidth]{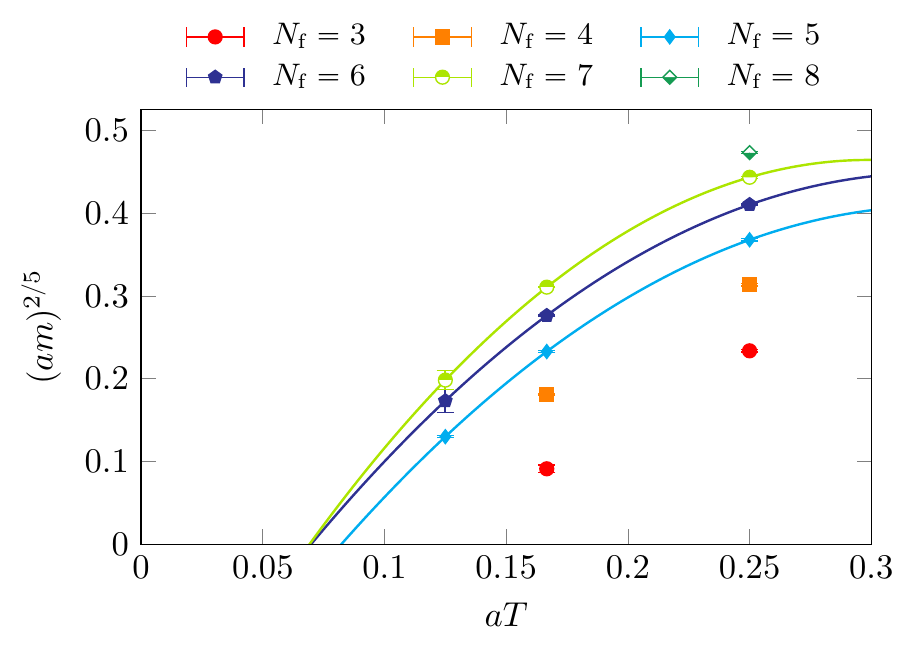}}
    \caption{%
      The chiral critical surface projected onto different planes.
      Every point represents a phase boundary with an implicitly tuned $\betac(am,\Nf,\NTau)$.
      Lines in \subref{fig:m_nf-fixed-nt} represent next-to-leading order scaling fits to \cref{eq:mscale} as in \cref{tab:mfit}.
      Solid lines in \subref{fig:m_nt-fixed-nf} are the analogue of the solid line in \cref{fig:m_nt-second}; refer to \cref{sec:m_nt} for more details.
    }
\end{figure}

\Cref{fig:m_nf-fixed-nt} shows the chiral critical surface projected into the plane of bare quark mass and number of flavours for different values of $\NTau$.
This is the lattice version of \cref{fig:columbia-plot-m-nf}.
Part of the data for $\NTau=4$~\cite{Cuteri:2017gci} and $\NTau=6$~\cite{Cuteri:2018wci} have been published in similar form previously, here we have added several $\Nf$ on $\NTau\in\{4,6\}$ and all of the $\NTau=8$ data.\footnote{Note that, for the present analysis, the result for $\Nf=3$ at $\NTau=6$ is simulated rather than borrowed from~\onlinecite{deForcrand:2007rq}.}
The parameter region below the curves corresponds to first-order transitions and above to analytic crossovers.
The $\Nf$-axis represents a triple line for larger values, whose latent heat decreases with diminishing $\Nf$ until it vanishes in a tricritical point $\Nf^\tric$.
The critical boundary line must therefore approach this point with scaling behaviour
\begin{equation}
    \Nf^c\big(am(\NTau),\NTau\big)= \Nf^\tric(\NTau)+\coeff{b}{1}(\NTau) (am)^{2/5} + \coeff{b}{2}(\NTau)(am)^{4/5}+\order[\big]{(am)^{6/5}}\;.
\end{equation}
Since the statistical error of our calculations is on the critical quark mass while $\Nf$ is exact, we consider the inverted series instead,
\begin{align}\label{eq:mscale}
    a\mc\big(\Nf(\NTau),\NTau\big)
    &= \coeff{D}{1}(\NTau)\big( \Nf-\Nf^\tric(\NTau)\big)^{5/2}\notag\\
    &+ \coeff{D}{2}(\NTau)\big( \Nf-\Nf^\tric(\NTau)\big)^{7/2} + \order[\Big]{\big( \Nf-\Nf^\tric(\NTau)\big)^{9/2}}\;,
\end{align}
where trivially $am^\tric(\NTau)=0$ in all cases.
It was already demonstrated in \onlinecite{Cuteri:2017gci} that for $\NTau=4$ the two lowest mass points in the present work and the $\Nf=2$ result from \onlinecite{Bonati:2014kpa} are consistent with a leading-order scaling curve.
However, as \cref{fig:m_nf-fixed-nt} shows, the scaling window is rather narrow in this variable pairing, and even a next-to-leading order fit breaks off the data rather quickly, both for $\NTau\in\{4,6\}$.
The reason is that scaling gets quickly superseded by a nearly linear behaviour over a large range of $\Nf$, analogous to $\Tc(\Nf)$ observed in the chiral limit of the continuum theory, cf.~\cref{fig:t-m-nf-phase-diagram}.

The emerging qualitative picture is thus: With increasing $\NTau$, i.e.~decreasing lattice spacing, the chiral phase transition observed on the lattice at fixed $\Nf$ weakens, so that the first-order region gets narrower.
This is in line with all earlier observations discussed in the introduction.
Because of the smallness of the scaling region, it is difficult to determine the value of $\Nf^\tric$ quantitatively.
On $\NTau=4$, the $\Nf=2$ theory has a first-order chiral transition, in agreement with a result based on an extrapolation from imaginary chemical potentials~\cite{Bonati:2014kpa}.
On $\NTau=6$ the $\Nf^\tric$ value shifts towards larger values and for $\NTau=8$ there are not enough data yet to constrain the scaling.
Nevertheless, the size of the observed cutoff effects strongly suggests further movement of the critical line on its way to the continuum.
One question is whether the critical quark masses scaled by the temperature will settle at finite values, which we checked is not the case yet.
However, \cref{fig:m_nf-fixed-nt} raises the additional question whether the  intercept $\Nf^\tric$ slides further to the right and possibly beyond $\Nf=3$.
More data and larger $\NTau$ are required for a definite answer, but we can get a hint by analysing yet another variable pairing.

\begin{table}[t]
    \setlength{\tabcolsep}{3mm}
    \renewcommand{\arraystretch}{1.1}
    \centering
    \begin{tabular}{
        @{\tabEdge}
        cc
        @{\hspace{8mm}}
        S[table-format=1.3(2)]S[table-format=1.4(2)]S[table-format=+1.4(1)]
        @{\hspace{8mm}}
        S[table-format=1.2, round-mode = places, round-precision = 2]
        @{\tabEdge}
      }
      \toprule
      $\NTau$ & range in $\Nf$ & {$\Nf^\tric$} & {$\coeff{D}{1}$} & {$\coeff{D}{2}$} & {$\chidof$} \\ 
      \midrule
      4 & $[2,2.7]$ &  \coeffDcenterNtIVWithError & \coeffDoneNtIVWithError & \coeffDtwoNtIVWithError  & \fitMNfNtIVchidof \\
      6 & $[2,4.5]$ &  \coeffDcenterNtVIWithError & \coeffDoneNtVIWithError & \coeffDtwoNtVIWithError  & \fitMNfNtVIchidof \\
      \bottomrule
    \end{tabular}
    \caption{Fits of the data in \cref{fig:m_nf-fixed-nt} to \cref{eq:mscale}.}
    \label{tab:mfit}
\end{table}

\subsection{Plane of quark mass and temperature}\label{sec:m_nt}

We now turn to the $\plane{am}{aT=\NTau^{-1}}$-plane.
\Cref{fig:m_nt-fixed-nf} shows a sequence of curves for different fixed $\Nf$.
The region below the lines corresponds to first-order transitions, that above to analytic crossovers.
For this variable pairing, the near-linear $\Nf$-dependence through $\Lambda_\text{QCD}$ does not dominate the behaviour of the individual curves, but rather shows up in the spacing between them.
However, a tricritical point is not guaranteed to exist for any given value of $\Nf$, so we have to test for the functional behaviour.

\begin{figure}[t]
    \centering
    \subcaptionbox{Fit in a first-order scenario.\label{fig:m_nt-first}}{\includegraphics[width=0.495\textwidth]{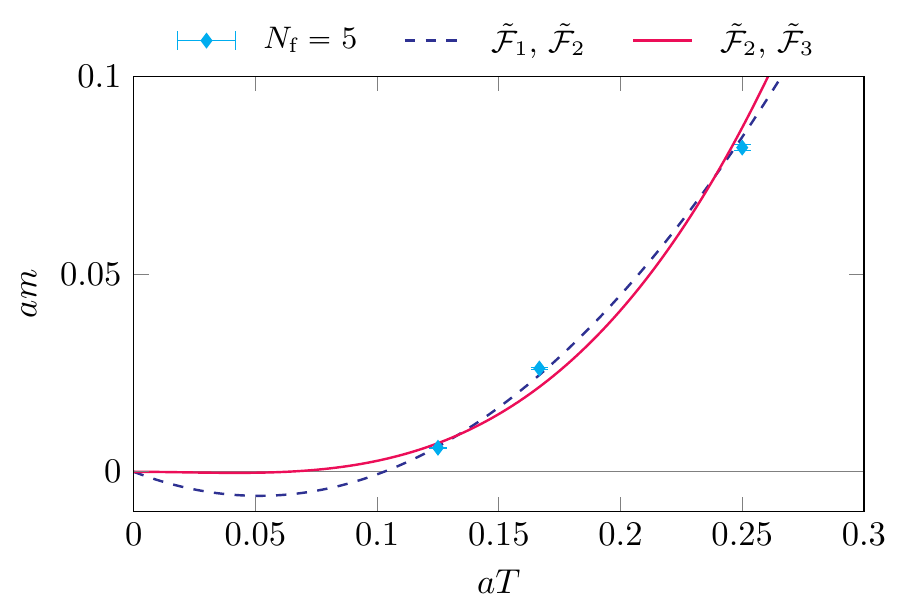}}\hfill
    \subcaptionbox{Fit in a second-order scenario.\label{fig:m_nt-second}}{\includegraphics[width=0.495\textwidth]{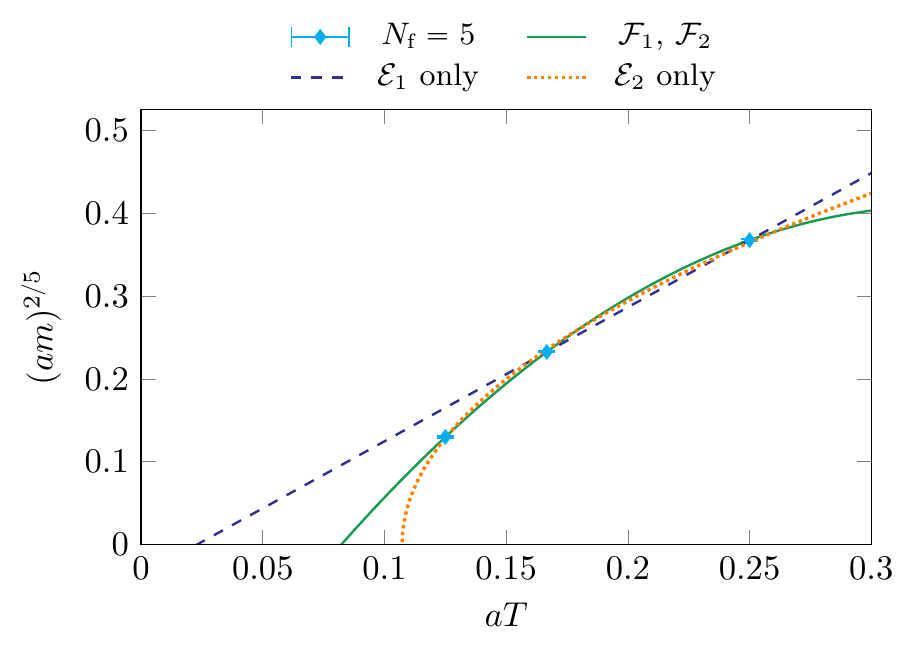}}
    \caption{%
      Chiral critical line for $\Nf=5$.
      Fits from \cref{tab:1st} pertaining to a first-order scenario, \cref{eq:m_T_1st}, do not describe the data.
      The data are instead consistent with a second-order scenario.
      In \subref{fig:m_nt-second} the green solid line interpolates the data according to \cref{eq:m_T} at truncated at next-to-leading order.
      Dashed and dotted lines represent interpolations according to \cref{eq:T_m_invertedLO,eq:T_m_invertedNLO}, respectively.
    }
    \label{fig:m_nt}
\end{figure}

Conventional wisdom predicts a definite first-order transition in the continuum for any $\Nf\geq 3$~\cite{Pisarski:1983ms,Butti:2003nu}.
This implies a finite $Z_2$-critical bare quark mass $\mc$ in physical units, with the usual $\order{a,a^2,\ldots}$ lattice corrections, and $a\mc(\NTau)\rightarrow 0$ in the limit $\NTau\rightarrow \infty$.
The critical lines in lattice units then approach the continuum as
\begin{equation}\label{eq:m_T_1st}
    a\mc(\NTau,\Nf)=\coeff{\tilde{F}}{1}(\Nf)\; aT+ \coeff{\tilde{F}}{2}(\Nf)\;(aT)^2 + \coeff{\tilde{F}}{3}(\Nf)\;(aT)^3 + \order[\big]{(aT)^4}\;.
\end{equation}
In \cref{fig:m_nt-first} and \cref{tab:1st} we fit our $\Nf=5$ data to two variations of this functional form.
Clearly our data are incompatible with \cref{eq:m_T_1st}, and the same holds for $\Nf\in\{6,7\}$.

The other possibility is for the critical lines to end on the $aT$-axis, which implies a tricritical point.
This is described by a polynomial in the scaling variable, which we invert as in the previous section,
\begingroup
\allowdisplaybreaks
\begin{align}
    a\Tc(am,\Nf)
    &= aT_\tric(\Nf) + \coeff{E}{1}(\Nf) (am)^{2/5} + \coeff{E}{2}(\Nf)(am)^{4/5} + \order[\big]{(am)^{6/5}}\;, \label{eq:T_m}\\[1ex]
    \Big(a\mc(\NTau,\Nf)\Big)^{2/5}
    &= \coeff{F}{1}(\Nf)\big(aT-aT_\tric(\Nf)\big) \notag \\
    &+ \coeff{F}{2}(\Nf)\big(aT-aT_\tric(\Nf)\big)^{2} + \order[\Big]{(aT-aT_\tric(\Nf)\big)^{3}}\;.\label{eq:m_T}
\end{align}
\endgroup
Unfortunately, with only three data points per line at most, meaningful next-to-leading order fits with non-vanishing intercept are impossible.
However, data in \cref{fig:m_nt-fixed-nf} with three $\NTau$-values show only a slight deviation from leading-order scaling, with an unambiguous curvature that can be covered by a next-to-leading order term as before.
In the following, we take this as an indication for tricritical scaling, for which a much stronger justification will be provided in \cref{sec:wilson_data}, and attempt to constrain the location of the tricritical points.

\begin{table}[t]
    \setlength{\tabcolsep}{5mm}
    \renewcommand{\arraystretch}{1.1}
    \centering
    \begin{tabular}{
        @{\tabEdge}
        c
        @{\hspace{8mm}}
        S[table-format=+1.3(1)]S[table-format=+1.3(2)]S[table-format=1.2(2)]
        @{\hspace{8mm}}
        S[table-format=3, round-mode = places, round-precision = 0]
        @{\tabEdge}
      }
      \toprule
      $\Nf$ & {$\coeff{\tilde{F}}{1}$} & {$\coeff{\tilde{F}}{2}$} & {$\coeff{\tilde{F}}{3}$} & {$\chidof$} \\
      \midrule
      \multirow{2}{*}{5}
      & \coeffFToneQuadraticNfVWithError & \coeffFTtwoQuadraticNfVWithError & {--} & \fitMTfirstQuadraticNfVchidof \\
      & {--} & \coeffFTtwoCubicNfVWithError & \coeffFTthreeCubicNfVWithError & \fitMTfirstCubicNfVchidof \\
      \midrule
      \multirow{2}{*}{6}
      & \coeffFToneQuadraticNfVIWithError & \coeffFTtwoQuadraticNfVIWithError & {--} & \fitMTfirstQuadraticNfVIchidof \\
      & {--} & \coeffFTtwoCubicNfVIWithError & \coeffFTthreeCubicNfVIWithError & \fitMTfirstCubicNfVIchidof \\
      \midrule
      \multirow{2}{*}{7}
      & \coeffFToneQuadraticNfVIIWithError & \coeffFTtwoQuadraticNfVIIWithError & {--} & \fitMTfirstQuadraticNfVIIchidof \\
      & {--} & \coeffFTtwoCubicNfVIIWithError & \coeffFTthreeCubicNfVIIWithError & \fitMTfirstCubicNfVIIchidof \\
      \bottomrule
    \end{tabular}
    \caption{Fits of the critical lines in \cref{fig:m_nt-first} to \cref{eq:m_T_1st}.}
    \label{tab:1st}
\end{table}

A next-to-leading order scaling interpolation, without error or $\chi^2$-estimate, can be obtained from \cref{eq:m_T} and is shown as the central line in \cref{fig:m_nt-second}.
We estimate the associated uncertainty to its $x$-axis crossing by alternatively interpolating our data with the inversion of \cref{eq:T_m} with a linear term only,
\begin{equation}\label{eq:T_m_invertedLO}
    \Big(a\mc(\NTau,\Nf)\Big)^{2/5} = \frac{1}{\coeff{E}{1}(\Nf)}\,\big(aT-aT_\tric(\Nf)\big)
\end{equation}
or fitting them to the inversion of \cref{eq:T_m} with a quadratic term only,
\begin{equation}\label{eq:T_m_invertedNLO}
    \Big(a\mc(\NTau,\Nf)\Big)^{2/5} = \sqrt{\frac{1}{\coeff{E}{2}(\Nf)}\,\big(aT-aT_\tric(\Nf)\big)}\;.
\end{equation}
The two $x$-axis crossings of \cref{eq:T_m_invertedLO,eq:T_m_invertedNLO} bracket the extrapolation using \cref{eq:m_T} at NLO.\footnote{Since the scaling field is $am$ and not $T$, \cref{eq:T_m} has to be preferred to \cref{eq:m_T} as starting point.}
\Cref{eq:T_m_invertedLO} is used to interpolate the data on the two coarsest lattices only, which corresponds to the most conservative choice.

While this procedure leaves a sizeable error band to be improved upon in the future, it still gives an interesting estimate of where the chiral critical surface ends in the massless limit.
In particular, the lower bounds on the tricritical values $aT_\tric(\Nf)$, or equivalently upper bounds on $\NTau^\tric(\Nf)$, can only be avoided by a change of sign in the curvatures of $(a\mc(\NTau))^{2/5}$ for $\NTau\geq 10$.

In the absence of such a drastic change, the near-tricritical scaling of our data suggests that for any fixed $\Nf \;\lesssim 6$ there is a maximal value $\NTau^\tric(\Nf)$, for which the chiral transition on the lattice can be of first order.
Following the qualitative behaviour in \cref{fig:m_nt-fixed-nf}, this value keeps increasing with $\Nf$.
There will then be a special (not necessarily integer) value $\Nf^0$, for which the tricritical point hits the origin of the figure, $aT_\tric(\Nf^0)=0$, and this is the situation when a tricritical point exists in the continuum.
Only for $\Nf>\Nf^0$ can a first-order phase transition persist all the way to the continuum.
This part of our analysis is consistent with $\Nf^0>6$.

\subsection{Plane of number of flavours and temperature}

\begin{figure}[t]
    \centering
    \includegraphics[width=0.67\textwidth]{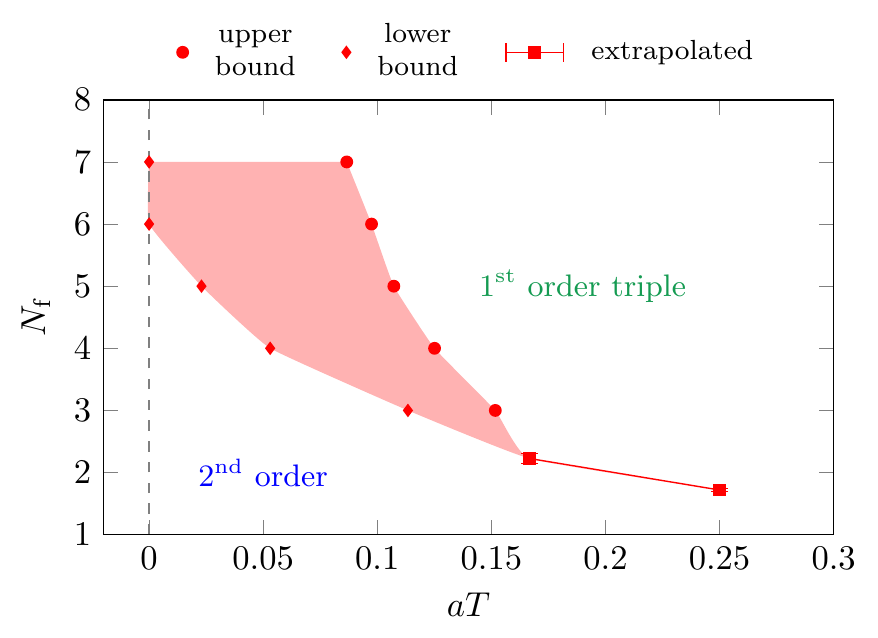}
    \caption{%
      Upper and lower bounds on the location of the tricritical line $\Nf^\tric(\NTau)$ for $am=0$.
      The lower left region represents second-order chiral transitions, the region above the band represents first-order transitions.
      These have no continuous connection to the continuum theory at $aT=0$ and are discretisation artefacts.
    }
    \label{fig:tricline}
\end{figure}

Finally one can also slice the chiral critical surface at fixed values of the quark mass $am$ and obtain the critical line separating first-order transitions and crossovers in the $\plane{\Nf}{\NTau^{-1}}$-plane, which reduces to a grid for integer variable values.
Such a diagram relies on the interpolation/extrapolation functions discussed in the previous sections.
The most interesting plane, and goal of our investigation, is the chiral limit, $am=0$, shown in \cref{fig:tricline}.
In this case the chiral phase transition must be non-analytic, and the tricritical ``line'' shown in the figure separates a region of first-order phase transitions above it from a region of second-order phase transitions below it.
The two rightmost points correspond to proper extrapolations from \cref{tab:mfit}, where the existence of the tricritical points is guaranteed and only the accuracy may vary.
By contrast, the error band to the left of it reflects the bounds estimated in the last subsection, and hinges upon better data description by tricritical scaling only.
We will come back to this issue in \cref{sec:wilson_data}.

Taking these chiral extrapolations at face value, it is intriguing to speculate how this tricritical line might continue beyond the $\Nf$-range covered by our simulations.
In fact, the entire first-order region corresponds to an area of triple points, which must be bounded by either a tricritical line or the boundaries of the $am=0$  parameter space.
Since our error band is again consistent with an approximately linear $\Nf$-dependence of $T_\tric(\Nf)$ on the lattice in the range $\Nf\in[2,7]$, and in the absence of any other special point, we conjecture the tricritical line to hit the $\Nf$-axis in the conformal point on the $\Nf$-axis, i.e., $\Nf^0=\Nf^*$.

\subsection{Implications for the chiral phase transition in the continuum}

Knowing the phase diagram in the bare parameter space of the lattice theory, we can draw conclusions for the continuum limit.
Here some care is in order.
A necessary condition to avoid problems with the rooting of the staggered fermion determinant requires the continuum limit to be taken before the chiral limit (for reviews of the staggered fermion discretisation and its potential problems, see \onlinecite{Sharpe:2006re,Kronfeld:2007ek} and the references therein).
A procedure to approach the continuum chiral transition for given $\Nf$ would then be:
\begin{enumerate}
    \item\label{task:I} Find the phase boundary $\betac(\NTau, am)$ for a chosen bare quark mass and $\NTau$;
    \item\label{task:II} Determine the order of the transition;
    \item\label{task:III} Set the scale for the lattice spacing, compute the pseudoscalar mass $\mPS(\betac,am)$ and the critical temperature $\Tc(am,\NTau)$;
    \item\label{task:IV} Repeat \cref{task:I,task:II,task:III} for a sequence of increasing $\NTau$ along a line of constant physics, i.e., tuning the bare quark mass such that  $\mPS(\betac,am)$ in physical units stays fixed;
    \item\label{task:V} Once $\NTau$ is large enough that the order of the transition does not change anymore, continuum extrapolation ($\beta\rightarrow \infty, \NTau^{-1}\rightarrow 0)$ gives $\Tc(\mPS)$;
    \item\label{task:VI} Repeat \cref{task:I,task:II,task:III,task:IV,task:V} for a sequence of decreasing pseudoscalar masses.
\end{enumerate}

\begin{figure}[t]
    \centering
    \subcaptionbox{First-order continuum transition.\label{fig:lcp-first}}{\includegraphics[width=0.5\textwidth, clip, trim=3mm 0 0 0]{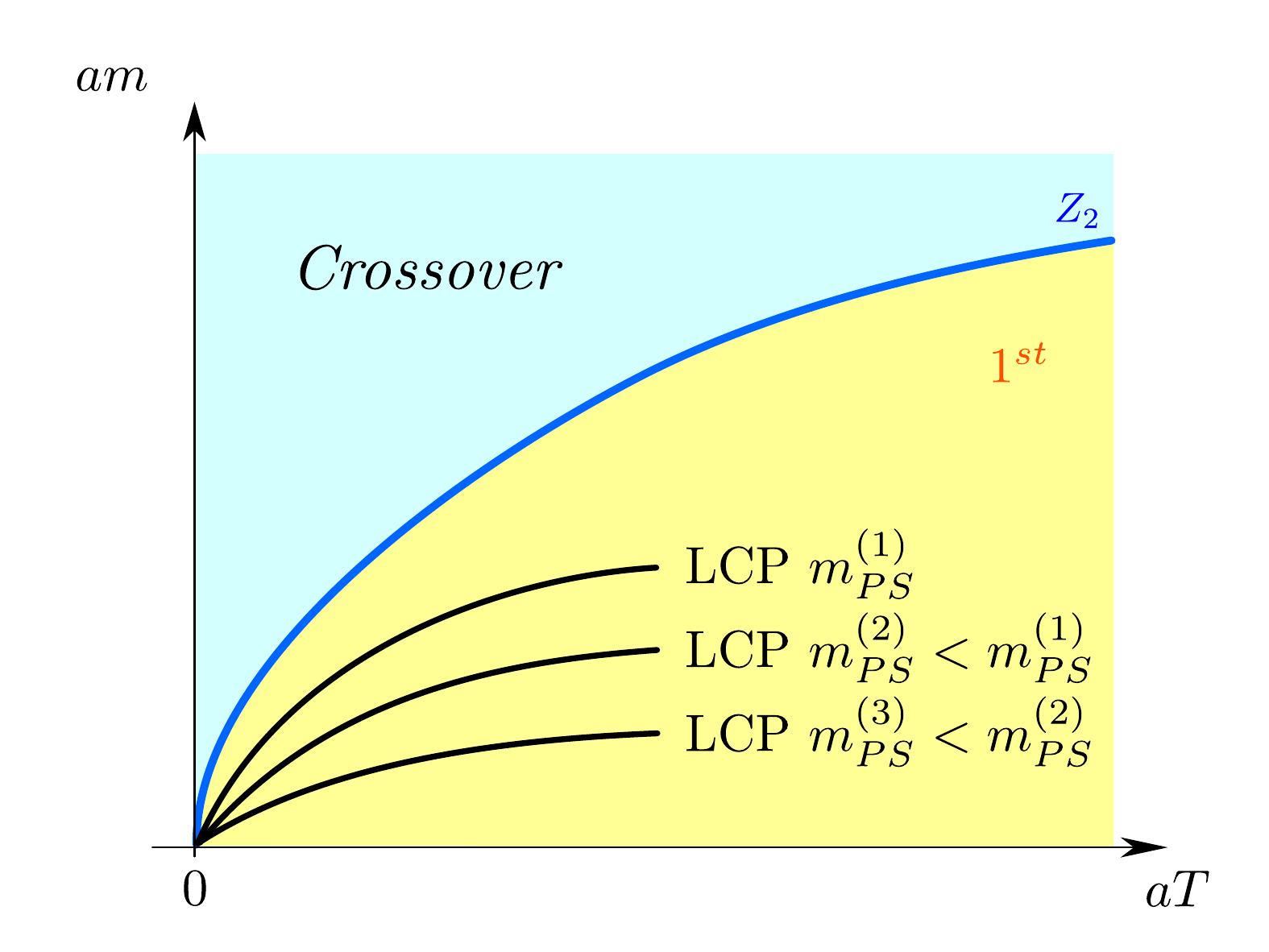}}\hfill
    \subcaptionbox{Second-order continuum transition.\label{fig:lcp-second}}{\includegraphics[width=0.5\textwidth, clip, trim=0 0 3mm 0]{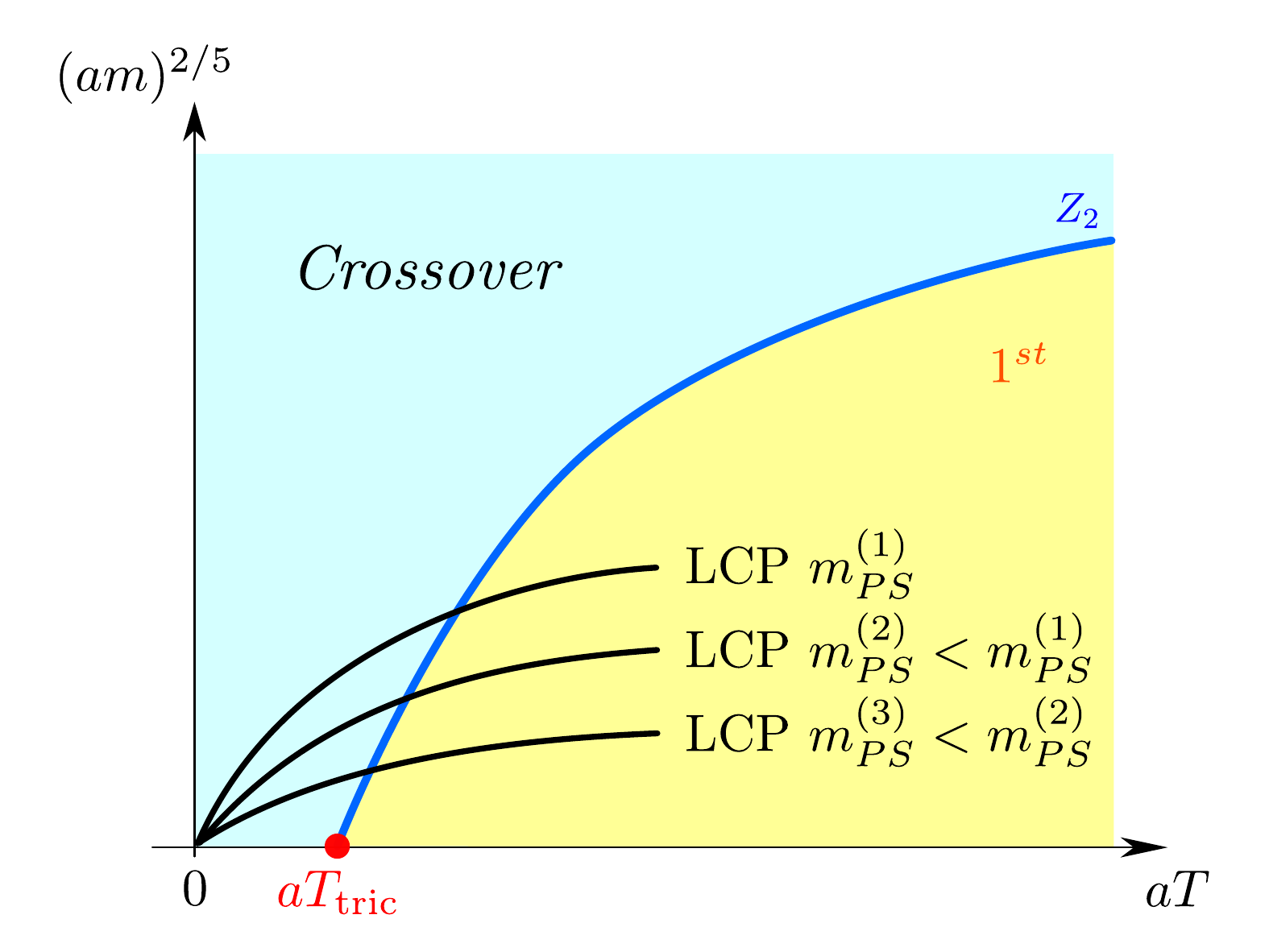}}
    \caption{%
      Different scenarios of the approach to the continuum chiral limit along several lines of constant physics.
    }
    \label{fig:lcp}
\end{figure}

The implications of our results are easiest to see in the $\plane{am}{aT}$ plots and illustrated schematically in \cref{fig:lcp}, where the continuum limit is located in the lower left corner, $am=aT=0$.
Continuum approaches for different physical masses $\mPS$ correspond to different lines of constant physics ending in the origin.
For every $\Nf$ with a finite $\NTau^\tric(\Nf)$, simulations with $\NTau>\NTau^\tric$ will show exclusively crossover behaviour for \textit{any} tuning of the bare masses $am(\beta)\neq 0$, \cref{fig:lcp-second}.
Consequently, for these $\Nf$ the chiral limit in the continuum corresponds to an isolated point with a second-order phase transition.
By contrast, a first-order chiral phase transition in the continuum extends over a finite mass range to a $Z_2$-point at some critical quark mass $\mc$, such as in the scenario \cref{fig:columbia-first}.
This requires the chiral critical line to terminate in the origin as $\NTau\rightarrow \infty$, cf.~\cref{fig:lcp-first}.
In such a case, there is no tricritical scaling and one expects a different functional form of the critical line.
Our data for $\Nf\in[2,6]$ clearly prefer the first option.

Of course, any conclusion has to be consistent between all parameter pairings.
For example, in \cref{fig:beta_m-fix-nf} the continuum limit is on the upper left corner, \mbox{$(\beta=\infty, am=0)$}.
To obtain a continuum first-order transition requires the critical line for every $\Nf$ in question to bend upwards, so as to include the entire $\beta$-axis in the first-order region.
Finally, in \cref{fig:tricline}, a continuum limit with first-order transitions for $\Nf\geq 3$ requires the tricritical line to hit the $\Nf$-axis below $\Nf=3$.

Based on the available data only, we thus have to conclude that for $2\leq \Nf\leq 6$, and possibly $2\leq \Nf\leq \Nf^*$, the regions of first-order transitions displayed in the bare parameter space of the standard staggered lattice theory are not continuously connected to the chiral limit in the continuum, and hence represent lattice artefacts.
Note that this conclusion is independent of whether or not additional unphysical phases, like a spontaneously broken staggered shift symmetry \cite{Cheng:2011ic,Kotov:2021mgp}, exist in the current discretisation.
Such phases cannot be detected by the chiral condensate alone and represent lattice artefacts as well.
Like the first-order regions described here, the system enters such phases when the chiral limit is approached before the continuum limit, while they are avoided by the correct ordering of limits. 

Lattice artefacts are discretisation dependent, i.e., the location of the chiral critical surface investigated here will be different for different lattice actions.
This would explain the fact that no first-order transition region is seen in simulations with highly improved staggered actions for $\Nf=3$~\cite{Bazavov:2017xul}.
It would be interesting to confirm this by investigating how the chiral critical lines shift under tuneable improvement, such as an increasing amount of smearing steps in the action.
One would expect an effect similar to that of decreasing the lattice spacing, as was indeed observed in an earlier study with a stout-smeared staggered action for $\Nf=3$ \cite{Varnhorst:2015lea}.
In particular, also the tricritical boundary line in \cref{fig:tricline} should then shift upwards, while its left end should remain anchored at the conformal point.
This is because at a second-order point in the continuum the correlation length diverges, so that the UV-differences between different discretisations are suppressed and all actions with a correct continuum limit must reproduce that point.

\section{Wilson fermions}

In view of our surprising conclusions for $\Nf\geq 3$, and in order to rule out potential problems with the rooting procedure of staggered fermions, it is desirable to cross check our analysis in Wilson's discretisation scheme.
In that case, matters are complicated by the fact that the Wilson term breaks chiral symmetry completely at any finite lattice spacing.
As a consequence there is an additive renormalisation of the quark mass, and the naive chiral condensate is always non-zero.
This raises the question whether an analysis based on three-state coexistence and tricritical scaling is applicable at all.
In this section we suggest how this could be realised in the bare parameter space of Wilson fermions, and find compelling evidence for a tricritical point in already published $\Nf=3$ Wilson data~\cite{Kuramashi:2020meg}.

\subsection{Parameter space of Wilson fermions for fixed \texorpdfstring{$\Nf$}{Nf}}

\begin{figure}[t]
    \centering
    \subcaptionbox{Schematic bare parameter phase diagram.\label{fig:wilson-phase-diagram}}{\includegraphics[width=0.49\textwidth]{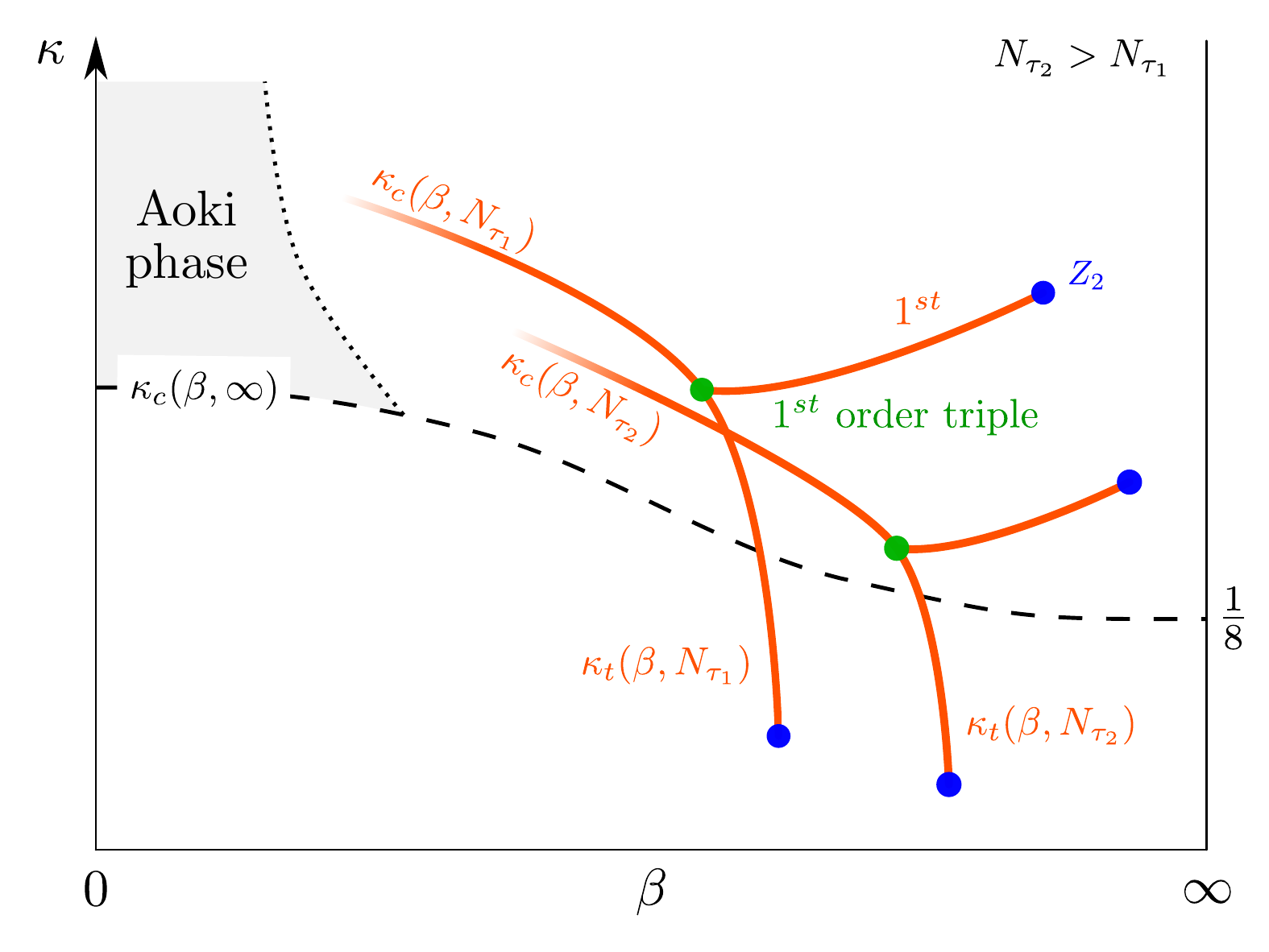}}\hfill
    \subcaptionbox{Columbia-like plot for Wilson fermions.\label{fig:wilson-columbia-like}}{\includegraphics[width=0.49\textwidth]{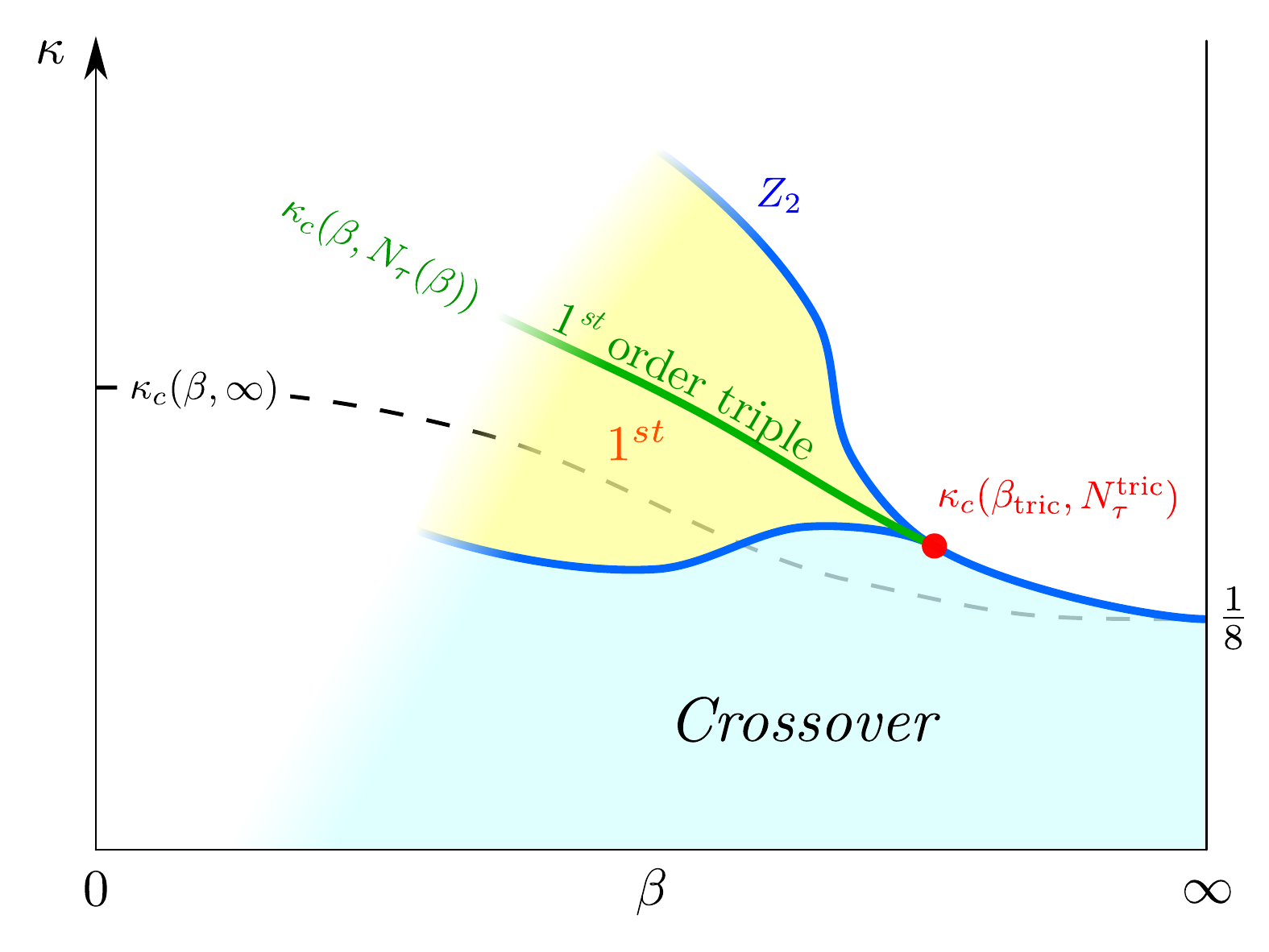}}
    \caption{%
      Suggested phase structure of finite temperature lattice QCD with Wilson fermions.
      In \subref{fig:wilson-columbia-like} $\NTau$ is implicitly tuned so that each point $(\kappa,\beta(\NTau))$ is on the phase boundary.
      The line $\kappac(\beta, \NTau(\beta))$ denotes the chiral limit at finite temperature and is a triple line, which ends in a tricritical point.
    }
    \label{fig:wilson}
\end{figure}

In the standard formulation of Wilson's fermion action with $\Nf$ degenerate flavours, the bare quark mass $am$ is traded for the hopping parameter,
\begin{equation}
    \kappa=\frac{1}{2(am+4)}\;.
\end{equation}
For the tree-level action, $\kappac=1/8$ represents a ``chiral limit'' defined by a vanishing bare quark mass.
In the interacting theory, this critical parameter value gets shifted as a function of the gauge coupling $\beta$, leading to an additive renormalisation of the quark mass,
\begin{equation}
    am_q=\frac{1}{2\kappa}-\frac{1}{2\kappac(\beta)}\;.
\end{equation}
A subtracted and renormalised chiral condensate can be defined based on an axial Ward-Takahashi identity~\cite{Bochicchio:1985xa},
\begin{equation}
    \chiralcond_R =2(am_q) Z \sum_x\langle \pi(x)\pi(0)\rangle\;,
\end{equation}
with $\pi(x)$ an interpolating operator for the pseudoscalar meson and a renormalisation factor $Z$.
Approaching the chiral limit, the pseudoscalar meson mass and the quark mass are related as in the continuum,
\begin{equation}\label{eq:chiral}
    a\mPS^2\;\propto\; am_q\;.
\end{equation} 
It is therefore customary to define $\kappac(\beta)$ by the vanishing of the pseudoscalar meson mass in the vacuum, i.e., $a\mPS(\kappac(\beta),\beta)=0$ at $\NTau=\infty$.
This is shown schematically as a dashed line in \cref{fig:wilson-phase-diagram}.
Towards the strong coupling region, this line meets the parity-flavour violating Aoki phase~\cite{Aoki:1983qi,Aoki:1986xr}, which ends in a cusp~\cite{Bitar:1996kc,Ilgenfritz:2003gw} whose location depends on the lattice action and the value of $\NTau$.
Around $\kappac(\beta)$, Wilson chiral perturbation for the theory predicts a metastability region corresponding to a first-order bulk transition between positive and negative quark mass, while the meson mass stays finite everywhere, both for untwisted and twisted mass~\cite{Sharpe:1998xm,Munster:2004am}.
A metastability region has been identified numerically at zero temperature~\cite{Farchioni:2004us} as well as at finite temperature~\cite{Blum:1994eh,Ilgenfritz:2009ns}, but its location and extent depend strongly on the chosen action and $\NTau$~\cite{Aoki:2004iq}.

\begin{figure}[t]
    \centering
    \includegraphics[width=0.6\textwidth]{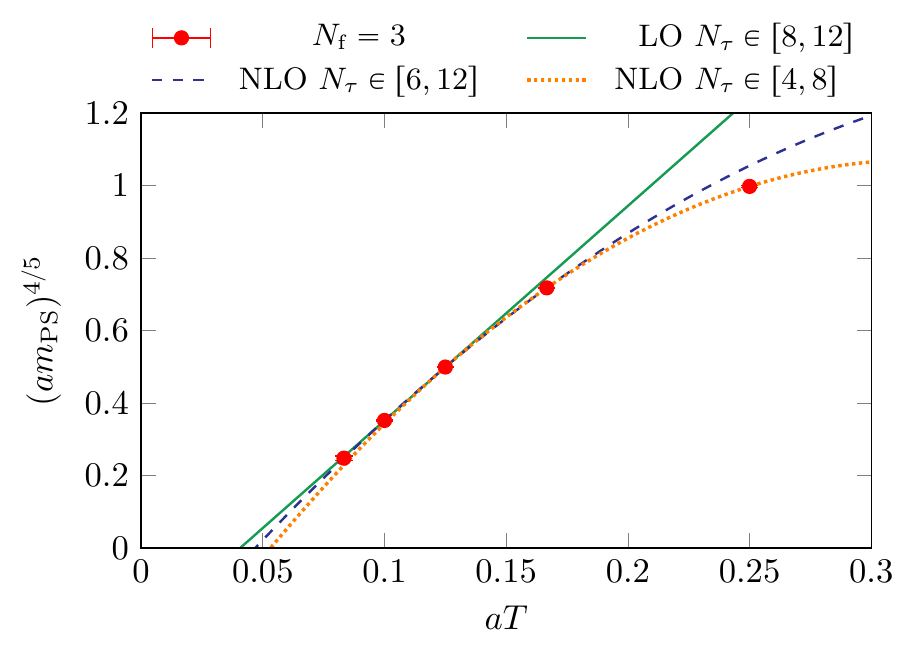}
    \caption{%
      Tricritical scaling of the critical pseudoscalar mass for $\Nf=3$ clover-improved Wilson fermions.
      The data are taken from Table IV in \protect\onlinecite{Kuramashi:2020meg}.
    }
    \label{fig:wilson_data}
\end{figure}

The series of $\Nf=3$ data~\cite{Jin:2014hea,Jin:2017jjp,Kuramashi:2020meg}, which we re-analyse below, is based on the RG-improved Iwasaki gauge action~\cite{Iwasaki:2011np} and a non-perturbatively $\order{a}$-improved Wilson clover fermion action~\cite{Aoki:2005et}.
We are not aware of a dedicated study of the bare phase diagram pertaining to the precise action and parameter tunings used in those simulations, besides determining the line  $\kappac(\beta,\NTau=\infty)$.
However, a previous study using the same action with a mean field tuning of the clover coefficient~\cite{AliKhan:2000wou} reports a phase diagram as sketched by the dashed lines in \cref{fig:wilson}, with no additional structures besides an Aoki phase in the strong coupling region, so we will base our discussion on this situation.

First, it has to be emphasised that for studies of the thermal phase transition we need the lines $\kappac(\beta,\NTau)$ for the finite $\NTau$ under consideration, and not $\kappac(\beta,\NTau=\infty)$, which is only needed to set the scale.
The former marks the vanishing of the pseudoscalar screening mass in the low temperature phase, and is related to the latter by an expansion in powers of $\NTau^{-1}=aT$,
\begin{equation}
    \kappac(\beta,\NTau)=\kappac(\beta,\infty)+\coeff{G}{1}(\beta)\,\NTau^{-1}+\coeff{G}{2}(\beta)\,\NTau^{-2} + \order[\big]{\NTau^{-3}}\;.
\end{equation}
In the literature the difference between the two is often dismissed, being of $\order{a}$, whereas in fact it is qualitatively crucial.
The partition function at finite $\NTau$ has no singularities on the line $\kappac(\beta,\infty)$ (except at its crossings with the thermal transition).
Furthermore, the subtracted chiral condensate has finite values with different signs across $\kappac(\beta,\NTau)$, which should therefore mark a first-order transition.\footnote{For the order of this transition it is immaterial whether the pseudoscalar screening mass is actually zero on the line, or whether it jumps between finite values.}
Following this line with increasing $\beta$ at fixed $\NTau$, the thermal chiral phase transition is reached at some critical coupling.
From this point the thermal transition lines $\kappa_t(\beta,\NTau)$ branch off into the positve and negative quark mass directions, respectively, along which the chiral transition weakens to end in a critical point.
At the branching point the line $\kappac(\beta,\NTau)$ should terminate, since on the large-$\beta$-side of the thermal transition the Matsubara modes $\sim 2\pi T$ produce an always non-zero screening mass and the subtracted chiral condensate can pass through zero smoothly.
The branching point $\kappac(\beta,\NTau)=\kappa_t(\beta,\NTau)$ then represents the Wilson version of the continuum triple point discussed in \cref{fig:pbp-3-state}.

For a sequence of increasing $\NTau[1]<\NTau[2]<\ldots$, the lines $\kappac(\beta,\NTau)$ gradually approach $\kappac(\beta,\infty)$, the triple points trace out a triple line and the critical endpoints form a chiral critical line separating first-order transitions for light quarks from the crossover region.
This is shown in the Columbia-like \cref{fig:wilson-columbia-like}, which gives the analogue of the critical quark mass lines at fixed $\Nf$ in the staggered case, \cref{fig:beta_m-fix-nf}.
There are then two possibilities for the situation at larger $\beta$-values: Either the triple line extends all the way to the continuum limit, or it terminates in a tricritical point, where it changes to a second-order transition, as sketched in \cref{fig:wilson-columbia-like}.

\subsection{Numerical analysis for \texorpdfstring{$\Nf=3$}{Nf}}\label{sec:wilson_data}

\begin{table}[t]
    \setlength{\tabcolsep}{5mm}
    \renewcommand{\arraystretch}{1.1}
    \centering
    \begin{tabular}{
        @{\tabEdge}
        c
        @{\hspace{8mm}}
        S[table-format=1.4(2)]S[table-format=1.2(2)]S[table-format=+2.1(2)]
        @{\hspace{8mm}}
        S[table-format=1.2, round-mode = places, round-precision = 2]
        @{\tabEdge}
      }
      \toprule
      range in $\NTau$ & {$aT_\tric$} & {$\coeff{H}{1}$} & {$\coeff{H}{2}$} & {$\chidof$} \\
      \midrule
      $[8,12]$ & \coeffHcenterNtVIIItoXIIWithError  &  \coeffHoneNtVIIItoXIIWithError & {--}                          & \fitMpsTNtVIIItoXIIchidof \\
      $[6,12]$ & \coeffHcenterNtVItoXIIWithError    &  \coeffHoneNtVItoXIIWithError   & {$\phantom{1}\coeffHtwoNtVItoXIIWithError$}  & \fitMpsTNtVItoXIIchidof   \\
      $[4,12]$ & \coeffHcenterNtIVtoXIIWithError    &  \coeffHoneNtIVtoXIIWithError   & \coeffHtwoNtIVtoXIIWithError  & \fitMpsTNtIVtoXIIchidof   \\
      $[4,8]$  & \coeffHcenterNtIVtoVIIIWithError   &  \coeffHoneNtIVtoVIIIWithError  & \coeffHtwoNtIVtoVIIIWithError & {--} \\
      \bottomrule
    \end{tabular}
    \caption{Fits of the data in \cref{fig:wilson_data} to \cref{eq:m_wilson}.}
    \label{tab:wilson_fits}
\end{table}

We now apply our analysis to the sequence of $\Nf=3$ clover-improved Wilson fermion simulations with $\NTau\in\{4,6,8,10,12\}$, whose data are published in \onlinecite{Jin:2014hea,Jin:2017jjp,Kuramashi:2020meg}.
In these works, the pseudoscalar meson mass was evaluated along the chiral critical endpoints in the lattice parameter space, which we reproduce in \cref{fig:wilson_data}.
The region below the critical line corresponds to first-order transitions, the region above to crossover.
To elucidate tricritical scaling with the quark mass, we employ \cref{eq:chiral} and convert the scaling relation for the quark mass as a function of $aT=\NTau^{-1}$, \cref{eq:m_T}, to a scaling relation for the pseudoscalar mass,
\begin{align}\label{eq:m_wilson}
    \big(a\mPSc(\NTau,\Nf)\big)^{4/5}
    &= \coeff{H}{1}(\Nf)\big(aT-aT_\tric(\Nf)\big) \notag\\
    &+ \coeff{H}{2}(\Nf)\big(aT-aT_\tric(\Nf)\big)^{2} + \order[\Big]{\big(aT-aT_\tric(\Nf)\big)^{3}}\;.
\end{align}
\Cref{fig:wilson_data} shows three different fits of the data to this scaling form, as detailed in \cref{tab:wilson_fits}.
The first fit demonstrates that the data for the smallest critical masses, corresponding to the finest $\NTau\in\{8,10,12\}$ lattices, are excellently described by leading-order tricritical scaling.
By contrast, they are incompatible with $a\mPSc\sim \NTau^{-1}\; (\chidof>100)$ and in tension with $a\mPSc\sim\NTau^{-2} \;(\chidof>3)$, one of which would be the expected behaviour in the absence of a tricritical point.
Moreover, a next-to-leading-order scaling fit including $\NTau=6$ is too good to be properly constrained, while fitting the entire range $\NTau\in[4,12]$ produces tension but only mild shifts the extracted parameters.
Finally, interpolating the next-to-leading order scaling form to $\NTau\in\{4,6,8\}$ data only, as we did in our staggered analysis in \cref{sec:m_nt}, $aT_\tric$ is only $\sim 25$\% off its value from the best fit.

Altogether the available Wilson data thus show:
\begin{enumerate}
    \item a clear preference for tricritical scaling and
    \item the stability of the analysis when restricted to the coarser lattices $\NTau\in\{4,6,8\}$, which we employed in \cref{sec:m_nt}.
\end{enumerate}
Avoiding the existence of a finite $\NTau^\tric$ would require a change of curvature in \cref{fig:wilson_data}  for $\NTau> 12$.
In the absence of such a drastic change, we conclude that any sequence of simulations with $\NTau>\NTau^\tric$ with $\NTau^\tric=(aT_\tric)^{-1}\approx 24$ will inevitably approach the continuum chiral limit from the crossover region, implying a second-order transition, in full accord with our staggered results.
Conversely this would imply, however, that simulations of the phase transition with $\NTau<24$ cannot reproduce the correct continuum physics despite the improvement of the action.
This has repercussions for, e.g., studies of the $U(1)_A$-anomaly with the Wilson action~\cite{Brandt:2016daq,Brandt:2019ksy}.
By contrast, unimproved staggered fermions should reproduce the second order of the $\Nf=3$ transition for $\NTau\gtrsim 10$ already.

\section{Conclusions}

\begin{figure}[t]
    \centering
    \savebox{\largestimage}{\includegraphics[width=0.48\textwidth]{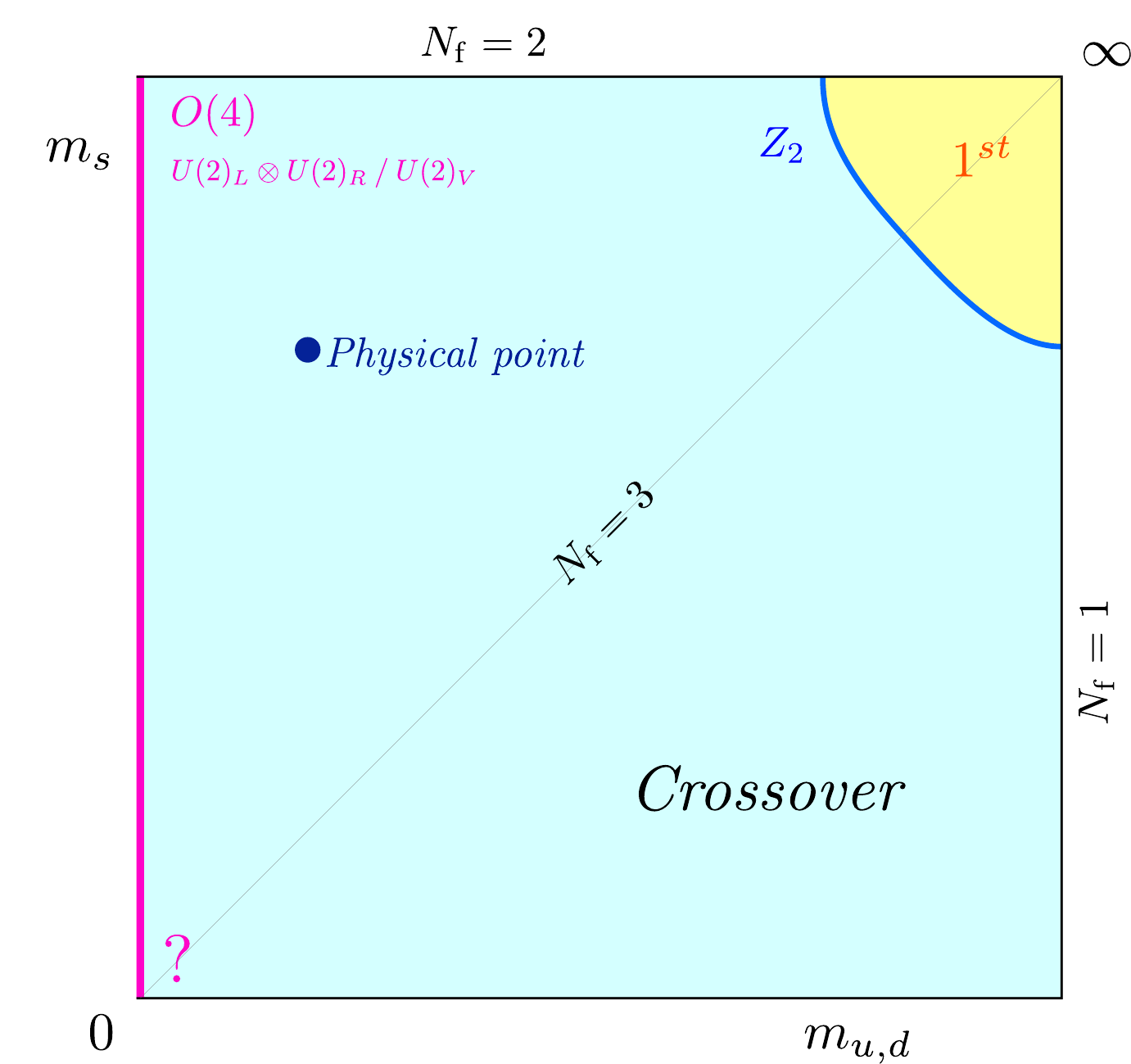}}%
        \setlength{\columnsep}{0.04\textwidth}
    \begin{floatrow}
        \ffigbox[0.48\textwidth][][b]{%
            \caption{%
                The Columbia plot in the continuum, as \mbox{predicted} by our analysis.
            }\label{fig:continuum-columbia-plot}
        }{%
            \usebox{\largestimage}
        }
        \ffigbox[0.48\textwidth][][b]{%
            \caption{%
                Suggested phase diagram for the chiral phase transition as a function of $\Nf$.
            }\label{fig:t-m-nf-phase-diagram-continuum}
        }{%
            \raisebox{\dimexpr.5\ht\largestimage-.5\height}{%
                \includegraphics[width=0.48\textwidth]{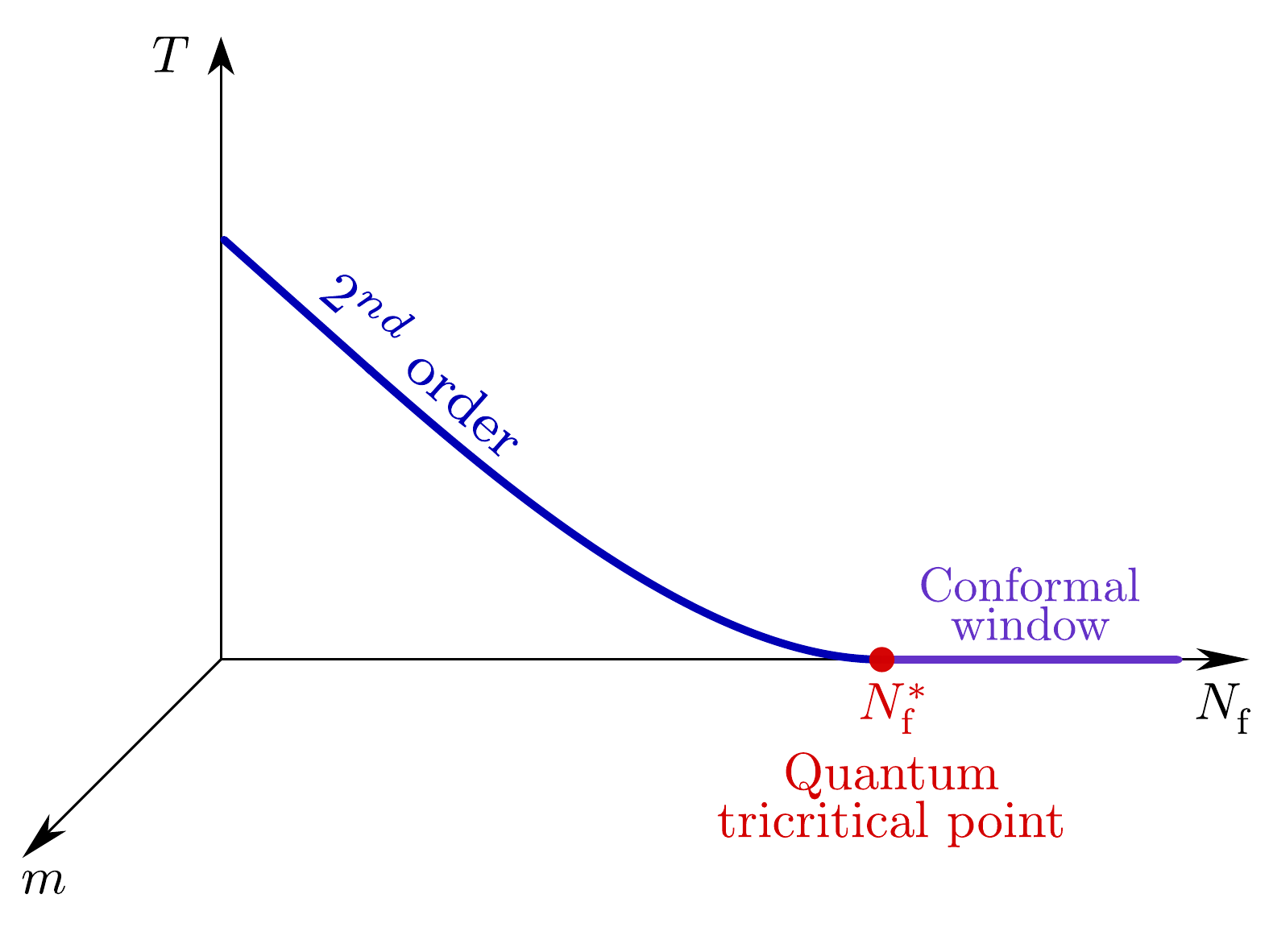}}
        }
    \end{floatrow}
\end{figure}

In summary, we have conducted a comprehensive analysis of the finite temperature chiral transitions observed in the bare parameter space $\{\beta,am,\Nf,\NTau\}$ of lattice QCD with unimproved staggered fermions.
In particular, we have mapped out the chiral critical surface, which separates the first-order transition region from the crossover region.
In the plane of vanishing bare quark mass, this surface terminates in a tricritical line $\NTau^\tric(\Nf)$, which for our presently available data is consistent with  $\NTau^\tric<\infty$ for all $\Nf\in[2,6]$.

The necessity to take the continuum limit before the chiral limit then enforces any appropriate series of simulations to approach the combined limits from the crossover region of the bare phase diagram.
This implies the chiral phase transition in the massless limit to be of second order for all $\Nf\leq 6$, and possibly up to the conformal window $\Nf\leq \Nf^*$.
As a crosscheck for our findings, we have reanalysed already published data from simulations with $\Nf=3$ $\order{a}$-improved Wilson fermions, which are equally consistent with tricritical scaling and a tricritical point at a finite $\NTau^\tric$.
Hence, this entirely different discretisation is consistent with the continuum chiral phase transition to be of second order as well.

Taking our results seriously, the continuum Columbia plot for $\Nf=2+1$ would have to be modified to look as in \cref{fig:continuum-columbia-plot}, with a second-order line all along the $\ms$-axis.
Our analysis has nothing to say about the universality class of this second-order line.
However, chiral symmetry being different in the limiting $\Nf\in\{2,3\}$ cases, one would expect the set of critical exponents associated with these transitions to smoothly cross from one universality class to the other as the strange quark mass varies.
Our findings for larger $\Nf$ suggest the phase diagram of many flavour QCD shown in \cref{fig:t-m-nf-phase-diagram-continuum}, with a second-order chiral phase transition in the massless limit, which ends in a quantum tricritical point at the onset of the conformal window.\footnote{A quantum tricritical point representing the zero temperature endpoint of a second order line is realised in the antiferromagnetic metal NbFe$_2$~\cite{Friedemann_2017}.}

Can our conclusions for $\Nf\geq 3$ be reconciled with the opposite predictions from 3D sigma-models~\cite{Pisarski:1983ms,Gausterer:1988fv,Butti:2003nu}?
These works investigate 3D effective Landau-Ginzburg-Wilson $\phi^4$ theories with the desired global symmetry and symmetry-breaking patterns, as well as a 'tHooft term added for the anomaly.
However, in mean field theory it is well known that $\phi^4$ theories cannot accommodate tricritical points, which requires a $\phi^6$-term.
A $\phi^6$-term is consistent with renormalisability in 3D, and indeed 3D $\phi^6$ theories display non-trivial infrared stable fixed points also in functional renormalisation group~\cite{Tetradis:1995br} and lattice~\cite{DelDebbio:2011zz} analyses.
Further studies of their connection to QCD would now be interesting.

While our data should be further supplemented for definitive conclusions concerning the continuum, an analysis employing tricritical scaling offers a new extrapolation tool for all discretisations observing first-order chiral transitions.

\acknowledgments

We thank J.~Braun, R.~Pisarski and D.~Rischke for a critical reading of an earlier version of the manuscript and many constructive remarks, as well as C.~Schmidt for discussions on the error analysis.
The authors acknowledge support by the Deutsche Forschungsgemeinschaft (DFG) through the grant CRC-TR 211 ``Strong-interaction matter under extreme conditions''.

\appendix
\section{Detailed overview of simulations}\label{sec:appendix}

In \cref{tab:simulation_overview_nt8,tab:simulation_overview_nt6,tab:simulation_overview_nt4}, a detailed overview of the accumulated statistics is offered.
Further information about previous simulations can be found in \onlinecite{Cuteri:2017gci}.

The total statistics refers to the whole spatial volume, and the statistics per chain can be inferred using the number of simulated $\beta$ values, considering that four Monte Carlo chains have been always run at each $\beta$ value.\footnote{Due to a few hardware failures, seven $\beta$ values across all data have been analysed with less than four chains, but always with more than two.}
Simulating multiple chains per fixed set of parameters helps in keeping under control the amount of accumulated statistics, and in particular in taking decisions about when to stop simulations.

The value of $\betac$ is reported without uncertainty and refers to the reweighted point which gives the closest skewness of the order parameter to zero.

The quantity $n_c(\Skewness)\approx 0$ denotes the number of chains at the outermost $\beta$ values (i.e.~the furthest $\beta$ from $\betac$), which are compatible with zero.
Ideally, this number should be 0, which would ensure the boundaries of the simulated $\beta$-range to belong to the two different phases of the system, while not being too far away from $\betac$.
This in turn would guarantee to be correctly cornering the phase transition within the given statistics.
Values different from zero require attention and can be accepted if small (e.g.~$1$) and rare.
Larger values signal a to-be-continued simulation, as it is still the case for several $\NTau=8$ sets.

Finally, $n_\sigma^\text{max}(\Skewness)$ denotes by how many standard deviations the skewness deviates between the two maximally separated chains.
This quantity should be as small as possible, and we usually aim at having it not much larger than 3.

\bigskip

\renewcommand{\P}{\phantom{1}}
\begin{table}[!hb]
    \setlength{\tabcolsep}{2mm}
    \newcommand{\CC}[1]{\cellcolor{#1}}
    \newcommand{\sep}{\,|\,}
    \renewcommand{\arraystretch}{1.1}
    \centering
    \begin{tabular}{cS[table-format=1.3]*{3}{r}}
        \toprule
        \multirow{2}{*}{$\Nf$} & {\multirow{2}{*}{$am$}} &  \multicolumn{3}{c}{$\betac$  | Total statistics | Number of $\beta$ values | $n_c(\Skewness)\approx 0$ | $n_\sigma^\text{max}(\Skewness)$ }\\
        &         &        \multicolumn{1}{c}{Aspect ratio $2$} & \multicolumn{1}{c}{Aspect ratio $3$} & \multicolumn{1}{c}{Aspect ratio $4$} \\
        \midrule
        \multirow{2}{*}{5.0}
        & 0.005  & 4.9549   \sep  426k\sep 3\sep 0\sep 2.6 & 4.95488  \sep  120k\sep 2\sep 0\sep 2.3 & \multicolumn{1}{c}{--} \\
        & 0.0075 & 4.9703   \sep  440k\sep 3\sep 0\sep 2.2 & 4.9706   \sep  189k\sep 3\sep 1\sep 2.6 & 4.9708   \sep   \P64k\sep 2\sep 4\sep 2.1 \\
        \midrule
        \multirow{3}{*}{6.0}
        & 0.01   & 4.8323   \sep  324k\sep 2\sep 0\sep 7.3 & 4.83198  \sep  404k\sep 4\sep 1\sep 4.0 & \multicolumn{1}{c}{--} \\
        & 0.0125 & \multicolumn{1}{c}{--}                  & 4.847    \sep  200k\sep 2\sep 0\sep 3.3 & \multicolumn{1}{c}{--} \\
        & 0.015  & 4.8611   \sep  360k\sep 2\sep 0\sep 1.2 & 4.8611   \sep  268k\sep 2\sep 0\sep 2.0 & 4.8618   \sep  168k\sep 2\sep 2\sep 1.9 \\
        \midrule
        \multirow{2}{*}{7.0}
        & 0.015  & 4.7164   \sep  440k\sep 3\sep 0\sep 2.1 & 4.71682  \sep  128k\sep 3\sep 3\sep 2.7 & \multicolumn{1}{c}{--} \\
        & 0.02   & 4.7444   \sep  480k\sep 3\sep 0\sep 4.2 & 4.745    \sep  260k\sep 3\sep 0\sep 3.0 & \multicolumn{1}{c}{--} \\
        \bottomrule
    \end{tabular}
    \caption{Statistics overview for $\NTau=8$.}
    \label{tab:simulation_overview_nt8}
\end{table}

\begin{table}[tb]
    \setlength{\tabcolsep}{2mm}
    \newcommand{\CC}[1]{\cellcolor{#1}}
    \newcommand{\sep}{\,|\,}
    \renewcommand{\arraystretch}{1.1}
    \centering
    \begin{tabular}{@{}lS[table-format=1.3]*{3}{r}@{}}
        \toprule
        \multirow{2}{*}{$\Nf$} & {\multirow{2}{*}{$am$}} &  \multicolumn{3}{c}{$\betac$  | Total statistics | Number of $\beta$ values | $n_c(\Skewness)\approx 0$ | $n_\sigma^\text{max}(\Skewness)$ }\\
        &         &        \multicolumn{1}{c}{Aspect ratio $2$} & \multicolumn{1}{c}{Aspect ratio $3$} & \multicolumn{1}{c}{Aspect ratio $4$} \\
        \midrule
        \multirow{2}{*}{3.0}
         & 0.0025 & 5.2122   \sep \P464k\sep 3\sep 0\sep 3.3 & 5.2126   \sep \P210k\sep 4\sep 0\sep 2.7 & \multicolumn{1}{c}{--} \\
         & 0.005  & 5.2218   \sep \P280k\sep 3\sep 0\sep 2.8 & 5.222    \sep \P300k\sep 3\sep 1\sep 2.5 & \multicolumn{1}{c}{--} \\
        \midrule
        \multirow{3}{*}{3.6}
         & 0.0075 & 5.1338   \sep \P532k\sep 4\sep 1\sep 3.7 & 5.13388  \sep \P433k\sep 3\sep 3\sep 2.7 & 5.1337   \sep 352k\sep 3\sep 0\sep 3.0 \\
         & 0.01   & 5.1427   \sep \P480k\sep 4\sep 0\sep 2.4 & 5.14275  \sep \P364k\sep 4\sep 0\sep 1.5 & 5.14294  \sep 360k\sep 3\sep 0\sep 1.2 \\
         & 0.0125 & 5.1519   \sep \P480k\sep 4\sep 0\sep 2.6 & 5.1519   \sep \P400k\sep 3\sep 0\sep 2.2 & 5.1519   \sep 480k\sep 4\sep 0\sep 1.2 \\
        \midrule
        \multirow{3}{*}{4.0}
         & 0.0125 & 5.0902   \sep \P640k\sep 4\sep 0\sep 2.4 & 5.09021  \sep \P536k\sep 3\sep 1\sep 4.3 & 5.09039  \sep 395k\sep 3\sep 2\sep 2.4 \\
         & 0.015  & 5.0991   \sep \P360k\sep 3\sep 0\sep 3.4 & 5.099    \sep \P644k\sep 5\sep 0\sep 2.5 & 5.09901  \sep 360k\sep 3\sep 0\sep 2.9 \\
         & 0.0165 & 5.1041   \sep \P351k\sep 3\sep 0\sep 2.4 & 5.1042   \sep \P360k\sep 3\sep 0\sep 3.8 & 5.1043   \sep 380k\sep 3\sep 0\sep 1.9 \\
        \midrule
        \multirow{3}{*}{4.4}
         & 0.015  & 5.0396   \sep \P553k\sep 3\sep 3\sep 3.7 & 5.03939  \sep \P526k\sep 4\sep 1\sep 4.2 & \multicolumn{1}{c}{--} \\
         & 0.02   & 5.0571   \sep \P360k\sep 3\sep 0\sep 1.5 & 5.0572   \sep \P460k\sep 3\sep 0\sep 1.6 & 5.05714  \sep 360k\sep 3\sep 1\sep 3.1 \\
         & 0.025  & 5.0738   \sep \P360k\sep 3\sep 0\sep 4.2 & 5.074    \sep \P360k\sep 3\sep 0\sep 2.2 & 5.0741   \sep 330k\sep 3\sep 0\sep 2.1 \\
        \midrule
        \multirow{3}{*}{5.0}
         & 0.02   & 4.9723   \sep \P400k\sep 3\sep 0\sep 4.8 & 4.97212  \sep \P439k\sep 3\sep 0\sep 7.6 & \multicolumn{1}{c}{--} \\
         & 0.025  & 4.9895   \sep \P480k\sep 3\sep 0\sep 2.6 & \textcolor{blue}{4.9893   \sep \P440k\sep 3\sep 0\sep 2.2} & 4.9893   \sep 235k\sep 3\sep 0\sep 1.6 \\
         & 0.03   & 5.0061   \sep \P360k\sep 3\sep 0\sep 1.9 & 5.0064   \sep \P360k\sep 3\sep 0\sep 2.0 & 5.0062   \sep 299k\sep 3\sep 0\sep 1.7 \\
        \midrule
        \multirow{3}{*}{6.0}
         & 0.035  & 4.8904   \sep \P920k\sep 3\sep 1\sep 2.3 & 4.89063  \sep \P978k\sep 3\sep 0\sep 9.4 & \multicolumn{1}{c}{--} \\
         & 0.04   & 4.9076   \sep \P920k\sep 4\sep 0\sep 2.1 & 4.9077   \sep \P760k\sep 3\sep 0\sep 2.7 & 4.90769  \sep 438k\sep 2\sep 2\sep 2.9 \\
         & 0.045  & 4.9243   \sep \P400k\sep 2\sep 2\sep 2.6 & 4.9245   \sep \P600k\sep 3\sep 0\sep 2.5 & 4.92444  \sep 612k\sep 3\sep 0\sep 3.8 \\
        \midrule
        \multirow{3}{*}{7.0}
         & 0.05   & 4.8181   \sep  1080k\sep 4\sep 2\sep 2.0 & 4.81816  \sep  1223k\sep 4\sep 1\sep 2.4 & 4.81838  \sep 560k\sep 3\sep 2\sep 2.6 \\
         & 0.055  & 4.835    \sep \P600k\sep 3\sep 0\sep 1.8 & 4.835    \sep \P559k\sep 3\sep 0\sep 2.0 & 4.83517  \sep 480k\sep 3\sep 0\sep 1.0 \\
         & 0.06   & 4.8517   \sep \P480k\sep 3\sep 0\sep 2.5 & 4.8517   \sep \P480k\sep 3\sep 0\sep 3.4 & 4.8518   \sep 322k\sep 3\sep 0\sep 2.5 \\
        \bottomrule
    \end{tabular}
    \caption{%
      Statistics overview for $\NTau=6$.
      The coloured value refers to $\NSigma=16$, which is slightly smaller than aspect ratio 3.
    }
    \label{tab:simulation_overview_nt6}
\end{table}

\clearpage
\newpage
\global\pdfpageattr\expandafter{\the\pdfpageattr/Rotate 90}
\begingroup

\begin{landscape}
    \setlength{\aboverulesep}{0pt}
    \setlength{\belowrulesep}{0pt}
    \setlength{\tabcolsep}{3mm}
    \newcommand{\CC}[1]{\cellcolor{#1}}
    \newcommand{\sep}{\,|\,}
    \renewcommand{\arraystretch}{1.1}
    \centering
    \begin{longtable}{cS[table-format=1.3]@{\hspace{1cm}}*{4}{r}}
        \toprule
        \multirow{2}{*}{$\Nf$} & {\multirow{2}{*}{$am$}} &  \multicolumn{4}{c}{$\betac$  | Total statistics | Number of $\beta$ values | $n_c(\Skewness)\approx 0$ | $n_\sigma^\text{max}(\Skewness)$ }\\
        &         &        \multicolumn{1}{c}{Aspect ratio $2$} & \multicolumn{1}{c}{Aspect ratio $3$} & \multicolumn{1}{c}{Aspect ratio $4$} & \multicolumn{1}{c}{Aspect ratio $5$} \\
        \midrule
        \endfirsthead
        \multicolumn{6}{l}{{\footnotesize\itshape continued from previous page}} \\
        \midrule
        \endhead
        \multicolumn{6}{r}{{\footnotesize\itshape continues on next page}} \\
        \endfoot
        \endlastfoot%
        \midrule
        \multirow{4}{*}{2.1}
         & 0.0015 & \multicolumn{1}{c}{--}                   & 5.2324   \sep  \P800k\sep 4\sep 0\sep 2.3 & 5.2323   \sep  \P505k\sep 4\sep 1\sep 4.6 & \multicolumn{1}{c}{--} \\
         & 0.0025 & \multicolumn{1}{c}{--}                   & 5.2345   \sep  \P804k\sep 3\sep 0\sep 2.1 & 5.23435  \sep  \P720k\sep 3\sep 0\sep 3.5 & 5.2343   \sep  \P260k\sep 3\sep 1\sep 2.7 \\
         & 0.0035 & \multicolumn{1}{c}{--}                   & 5.2366   \sep  \P818k\sep 3\sep 1\sep 4.6 & 5.2365   \sep  \P720k\sep 3\sep 0\sep 2.9 & 5.23647  \sep  \P380k\sep 3\sep 0\sep 2.2 \\
         & 0.0045 & \multicolumn{1}{c}{--}                   & 5.2386   \sep  \P720k\sep 3\sep 1\sep 3.7 & 5.2386   \sep  \P760k\sep 3\sep 0\sep 2.6 & 5.23855  \sep  \P460k\sep 3\sep 0\sep 2.5 \\
        \midrule
        \multirow{4}{*}{2.2}
         & 0.0025 & \multicolumn{1}{c}{--}                   & 5.2171   \sep  1175k\sep 5\sep 0\sep 3.4 & 5.2169   \sep \P760k\sep 3\sep 0\sep 2.3 & 5.21695  \sep \P660k\sep 3\sep 0\sep 2.5 \\
         & 0.005  & \multicolumn{1}{c}{--}                   & 5.2225   \sep \P840k\sep 3\sep 1\sep 2.1 & 5.2224   \sep  1080k\sep 4\sep 0\sep 2.2 & 5.22235  \sep  1035k\sep 4\sep 0\sep 2.8 \\
         & 0.0075 & \multicolumn{1}{c}{--}                   & 5.2275   \sep \P920k\sep 3\sep 0\sep 4.1 & 5.2276   \sep  1119k\sep 4\sep 0\sep 2.2 & 5.2274   \sep \P680k\sep 3\sep 0\sep 2.4 \\
         & 0.01   & \multicolumn{1}{c}{--}                   & 5.23255  \sep  2000k\sep 7\sep 0\sep 3.5 & 5.23235  \sep  1140k\sep 4\sep 0\sep 2.1 & 5.2323   \sep \P960k\sep 4\sep 0\sep 2.7 \\
        \midrule
        \multirow{4}{*}{2.4}
         & 0.0075 & 5.1942   \sep  1400k\sep 4\sep 0\sep 2.3 & 5.1939   \sep \P950k\sep 4\sep 0\sep 1.9 & 5.1938   \sep  1028k\sep 4\sep 0\sep 3.0 & \multicolumn{1}{c}{--} \\
         & 0.01   & 5.1993   \sep \P920k\sep 4\sep 0\sep 3.0 & 5.199    \sep  1081k\sep 3\sep 0\sep 2.6 & 5.19898  \sep  1050k\sep 3\sep 0\sep 1.8 & 5.19884  \sep  \P240k\sep 3\sep 1\sep 3.4 \\
         & 0.0125 & 5.2044   \sep  1121k\sep 4\sep 0\sep 2.9 & 5.2041   \sep \P960k\sep 3\sep 0\sep 1.9 & 5.20394  \sep \P963k\sep 3\sep 0\sep 2.3 & 5.2039   \sep  \P480k\sep 4\sep 0\sep 2.5 \\
         & 0.015  & 5.2091   \sep  1040k\sep 4\sep 0\sep 2.5 & 5.2088   \sep \P880k\sep 3\sep 0\sep 2.7 & 5.20853  \sep \P680k\sep 3\sep 0\sep 1.8 & 5.2087   \sep  \P360k\sep 3\sep 1\sep 2.5 \\
        \midrule
        \multirow{4}{*}{2.6}
         & 0.0125 & 5.1718   \sep  1080k\sep 4\sep 0\sep 3.5 & 5.1712   \sep \P920k\sep 3\sep 0\sep 2.3 & 5.1713   \sep  1480k\sep 4\sep 0\sep 3.0 & \multicolumn{1}{c}{--} \\
         & 0.015  & 5.1765   \sep  1240k\sep 3\sep 1\sep 2.8 & 5.1766   \sep  1280k\sep 4\sep 0\sep 3.4 & 5.17636  \sep  1080k\sep 3\sep 1\sep 1.9 & \multicolumn{1}{c}{--} \\
         & 0.0175 & 5.1816   \sep  1080k\sep 3\sep 0\sep 3.9 & 5.1814   \sep  1170k\sep 4\sep 0\sep 2.8 & 5.18135  \sep \P945k\sep 4\sep 0\sep 2.6 & \multicolumn{1}{c}{--} \\
         & 0.02   & 5.1868   \sep \P960k\sep 4\sep 0\sep 2.4 & 5.1861   \sep \P915k\sep 3\sep 0\sep 3.0 & 5.1861   \sep \P782k\sep 3\sep 0\sep 3.7 & \multicolumn{1}{c}{--} \\
        \midrule
        \multirow{4}{*}{2.8}
         & 0.0175 & 5.1503   \sep  1317k\sep 4\sep 0\sep 3.9 & 5.1499   \sep  1200k\sep 3\sep 0\sep 3.0 & 5.1498   \sep  1360k\sep 4\sep 0\sep 2.5 & \multicolumn{1}{c}{--} \\
         & 0.02   & 5.1553   \sep \P940k\sep 4\sep 0\sep 1.9 & 5.155    \sep  1160k\sep 3\sep 0\sep 3.4 & 5.1548   \sep \P556k\sep 3\sep 0\sep 2.1 & \multicolumn{1}{c}{--} \\
         & 0.0225 & 5.16     \sep  1400k\sep 4\sep 1\sep 2.1 & 5.15968  \sep \P790k\sep 3\sep 1\sep 2.4 & 5.15967  \sep  1060k\sep 3\sep 1\sep 3.8 & \multicolumn{1}{c}{--} \\
         & 0.025  & 5.1649   \sep  1120k\sep 4\sep 0\sep 2.5 & 5.1645   \sep \P840k\sep 3\sep 0\sep 3.1 & 5.16446  \sep \P866k\sep 3\sep 1\sep 4.7 & \multicolumn{1}{c}{--} \\
        \midrule
        \pagebreak
        \multirow{4}{*}{3.0}
         & 0.0225 & 5.1297   \sep  \P240k\sep 2\sep 5\sep 4.3 & 5.1291   \sep  300k\sep 3\sep 0\sep 1.6 & 5.129    \sep  \P270k\sep 3\sep 0\sep 2.4 & \multicolumn{1}{c}{--} \\
         & 0.025  & 5.1347   \sep  \P360k\sep 3\sep 0\sep 2.8 & 5.1341   \sep  415k\sep 3\sep 0\sep 1.6 & 5.1343   \sep  \P360k\sep 3\sep 0\sep 4.2 & \multicolumn{1}{c}{--} \\
         & 0.0275 & 5.1395   \sep  \P360k\sep 3\sep 0\sep 2.5 & 5.1391   \sep  480k\sep 4\sep 0\sep 2.8 & 5.139    \sep  \P360k\sep 3\sep 0\sep 3.8 & \multicolumn{1}{c}{--} \\
         & 0.03   & 5.1441   \sep  \P360k\sep 3\sep 1\sep 3.3 & 5.144    \sep  895k\sep 5\sep 0\sep 2.0 & 5.1438   \sep  \P400k\sep 3\sep 0\sep 2.4 & \multicolumn{1}{c}{--} \\
        \midrule
        \multirow{4}{*}{4.0}
         & 0.05   & 5.0431   \sep  \P360k\sep 3\sep 1\sep 2.3 & 5.0428   \sep  620k\sep 3\sep 0\sep 2.5 & 5.0429   \sep  \P560k\sep 3\sep 1\sep 4.9 & \multicolumn{1}{c}{--}  \\
         & 0.055  & 5.0538   \sep  \P400k\sep 3\sep 0\sep 3.4 & 5.0529   \sep  600k\sep 3\sep 0\sep 2.2 & 5.053    \sep  \P400k\sep 3\sep 3\sep 2.7 & \multicolumn{1}{c}{--}  \\
         & 0.06   & 5.0627   \sep  \P360k\sep 3\sep 1\sep 2.4 & 5.0625   \sep  440k\sep 3\sep 1\sep 2.6 & 5.0626   \sep  \P240k\sep 2\sep 1\sep 2.2 & \multicolumn{1}{c}{--}  \\
         & 0.065  & 5.0721   \sep  \P360k\sep 3\sep 0\sep 3.1 & 5.0721   \sep  360k\sep 3\sep 1\sep 3.0 & 5.072    \sep  \P360k\sep 3\sep 0\sep 1.8 & \multicolumn{1}{c}{--}  \\
        \midrule
        \multirow{5}{*}{5.0}
         & 0.07   & 4.9588   \sep  \P240k\sep 2\sep 2\sep 1.8 & 4.9579   \sep  400k\sep 3\sep 0\sep 2.1 & \multicolumn{1}{c}{--} & \multicolumn{1}{c}{--} \\
         & 0.075  & 4.9687   \sep  \P640k\sep 4\sep 0\sep 3.8 & 4.9688   \sep  359k\sep 3\sep 1\sep 1.7 & 4.9688   \sep  \P360k\sep 3\sep 0\sep 2.9 & \multicolumn{1}{c}{--} \\
         & 0.08   & 4.9793   \sep  \P560k\sep 4\sep 0\sep 2.7 & 4.9789   \sep  360k\sep 3\sep 1\sep 5.1 & 4.9791   \sep  \P400k\sep 3\sep 0\sep 2.4 & \multicolumn{1}{c}{--} \\
         & 0.085  & 4.9893   \sep  \P360k\sep 3\sep 1\sep 2.5 & 4.989    \sep  360k\sep 3\sep 0\sep 2.0 & 4.9888   \sep  \P360k\sep 3\sep 0\sep 2.5 & \multicolumn{1}{c}{--} \\
         & 0.09   & 4.999    \sep  \P480k\sep 4\sep 0\sep 2.2 & 4.9984   \sep  360k\sep 3\sep 0\sep 2.0 & 4.9985   \sep  \P240k\sep 2\sep 0\sep 4.5 & \multicolumn{1}{c}{--} \\
        \midrule
        \multirow{3}{*}{6.0}
         & 0.1    & 4.9075   \sep  1120k\sep 3\sep 1\sep 3.2 & 4.9077   \sep  920k\sep 2\sep 1\sep 3.2 & 4.90755  \sep  1360k\sep 3\sep 0\sep 4.4 & \multicolumn{1}{c}{--} \\
         & 0.11   & 4.9284   \sep \P960k\sep 3\sep 0\sep 3.6 & 4.928    \sep  600k\sep 3\sep 0\sep 2.2 & 4.9282   \sep \P760k\sep 3\sep 0\sep 3.8 & \multicolumn{1}{c}{--} \\
         & 0.12   & 4.9481   \sep \P600k\sep 3\sep 0\sep 2.0 & 4.948    \sep  600k\sep 3\sep 0\sep 2.3 & 4.948    \sep \P400k\sep 2\sep 1\sep 3.2 & \multicolumn{1}{c}{--} \\
        \midrule
        \multirow{3}{*}{7.0}
         & 0.12   & 4.8467   \sep  \P920k\sep 4\sep 0\sep 3.1 & 4.8465   \sep  960k\sep 3\sep 0\sep 2.3 & 4.8464   \sep  1160k\sep 3\sep 3\sep 4.1 & \multicolumn{1}{c}{--} \\
         & 0.13   & 4.8679   \sep  \P600k\sep 3\sep 0\sep 2.2 & 4.8677   \sep  760k\sep 3\sep 0\sep 2.9 & 4.8677   \sep \P640k\sep 3\sep 0\sep 2.0 & \multicolumn{1}{c}{--} \\
         & 0.14   & 4.888    \sep  \P600k\sep 3\sep 0\sep 2.7 & 4.88816  \sep  600k\sep 3\sep 1\sep 2.4 & 4.8883   \sep \P400k\sep 2\sep 2\sep 1.5 & \multicolumn{1}{c}{--} \\
        \midrule
        \multirow{3}{*}{8.0}
         & 0.14   & 4.7936   \sep   \P520k\sep 2\sep 0\sep 1.7 & 4.7938   \sep  320k\sep 2\sep 0\sep 1.5 & 4.7939   \sep  \P600k\sep 3\sep 0\sep 3.6 & \multicolumn{1}{c}{--} \\
         & 0.155  & 4.8265   \sep   \P400k\sep 3\sep 0\sep 2.7 & 4.8261   \sep  240k\sep 3\sep 0\sep 4.2 & 4.8261   \sep  \P160k\sep 2\sep 0\sep 2.5 & \multicolumn{1}{c}{--} \\
         & 0.17   & 4.8568   \sep  \P\P80k\sep 4\sep 0\sep 2.3 & 4.857    \sep  200k\sep 2\sep 0\sep 1.7 & 4.8571   \sep  \P240k\sep 2\sep 0\sep 1.9 & \multicolumn{1}{c}{--} \\
        \bottomrule
        \caption{Statistics overview for $\NTau=4$.}
        \label{tab:simulation_overview_nt4}
    \end{longtable}
\end{landscape}

\endgroup
\clearpage
\newpage
\global\pdfpageattr\expandafter{\the\pdfpageattr/Rotate 0}

\bibliographystyle{JHEP}
\bibliography{bibliography}

\providecommand{\href}[2]{#2}\begingroup\raggedright\begin{thebibliography}{10}

\bibitem{Aoki:2006we}
Y.~Aoki, G.~Endrodi, Z.~Fodor, S.D.~Katz and K.K.~Szabo, \emph{{The Order of
  the quantum chromodynamics transition predicted by the standard model of
  particle physics}}, \href{https://doi.org/10.1038/nature05120}{\emph{Nature}
  {\bfseries 443} (2006) 675}
  [\href{https://arxiv.org/abs/hep-lat/0611014}{{\ttfamily hep-lat/0611014}}].

\bibitem{Rajagopal:2000wf}
K.~Rajagopal and F.~Wilczek, \emph{{The Condensed matter physics of QCD}},  in
  \emph{At the frontier of particle physics. Handbook of QCD. Vol. 1-3},
  M.~Shifman and B.~Ioffe, eds., pp.~2061--2151 (2000),
  \href{https://doi.org/10.1142/9789812810458_0043}{DOI}
  [\href{https://arxiv.org/abs/hep-ph/0011333}{{\ttfamily hep-ph/0011333}}].

\bibitem{Brown:1990ev}
F.R.~Brown, F.P.~Butler, H.~Chen, N.H.~Christ, Z.-h.~Dong, W.~Schaffer et~al.,
  \emph{{On the existence of a phase transition for QCD with three light
  quarks}}, \href{https://doi.org/10.1103/PhysRevLett.65.2491}{\emph{Phys. Rev.
  Lett.} {\bfseries 65} (1990) 2491}.

\bibitem{Boyd:1996bx}
G.~Boyd, J.~Engels, F.~Karsch, E.~Laermann, C.~Legeland, M.~Lutgemeier et~al.,
  \emph{{Thermodynamics of SU(3) lattice gauge theory}},
  \href{https://doi.org/10.1016/0550-3213(96)00170-8}{\emph{Nucl. Phys. B}
  {\bfseries 469} (1996) 419}
  [\href{https://arxiv.org/abs/hep-lat/9602007}{{\ttfamily hep-lat/9602007}}].

\bibitem{Ejiri:2019csa}
{\scshape WHOT-QCD} collaboration, \emph{{End point of the first-order phase
  transition of QCD in the heavy quark region by reweighting from quenched
  QCD}}, \href{https://doi.org/10.1103/PhysRevD.101.054505}{\emph{Phys. Rev. D}
  {\bfseries 101} (2020) 054505}
  [\href{https://arxiv.org/abs/1912.10500}{{\ttfamily 1912.10500}}].

\bibitem{Cuteri:2020yke}
F.~Cuteri, O.~Philipsen, A.~Sch\"on and A.~Sciarra, \emph{{Deconfinement
  critical point of lattice QCD with $N_f$=2 Wilson fermions}},
  \href{https://doi.org/10.1103/PhysRevD.103.014513}{\emph{Phys. Rev. D}
  {\bfseries 103} (2021) 014513}
  [\href{https://arxiv.org/abs/2009.14033}{{\ttfamily 2009.14033}}].

\bibitem{Pisarski:1983ms}
R.D.~Pisarski and F.~Wilczek, \emph{{Remarks on the Chiral Phase Transition in
  Chromodynamics}}, \href{https://doi.org/10.1103/PhysRevD.29.338}{\emph{Phys.
  Rev.} {\bfseries D29} (1984) 338}.

\bibitem{Gausterer:1988fv}
H.~Gausterer and S.~Sanielevici, \emph{{Can the Chiral Transition in {QCD} Be
  Described by a Linear $\sigma$ Model in Three-dimensions?}},
  \href{https://doi.org/10.1016/0370-2693(88)91188-4}{\emph{Phys. Lett. B}
  {\bfseries 209} (1988) 533}.

\bibitem{Butti:2003nu}
A.~Butti, A.~Pelissetto and E.~Vicari, \emph{{On the nature of the finite
  temperature transition in QCD}},
  \href{https://doi.org/10.1088/1126-6708/2003/08/029}{\emph{JHEP} {\bfseries
  08} (2003) 029} [\href{https://arxiv.org/abs/hep-ph/0307036}{{\ttfamily
  hep-ph/0307036}}].

\bibitem{Pelissetto:2013hqa}
A.~Pelissetto and E.~Vicari, \emph{{Relevance of the axial anomaly at the
  finite-temperature chiral transition in QCD}},
  \href{https://doi.org/10.1103/PhysRevD.88.105018}{\emph{Phys. Rev. D}
  {\bfseries 88} (2013) 105018}
  [\href{https://arxiv.org/abs/1309.5446}{{\ttfamily 1309.5446}}].

\bibitem{Braun:2020ada}
J.~Braun, W.-j.~Fu, J.M.~Pawlowski, F.~Rennecke, D.~Rosenbl\"uh and S.~Yin,
  \emph{{Chiral susceptibility in ( 2+1 )-flavor QCD}},
  \href{https://doi.org/10.1103/PhysRevD.102.056010}{\emph{Phys. Rev. D}
  {\bfseries 102} (2020) 056010}
  [\href{https://arxiv.org/abs/2003.13112}{{\ttfamily 2003.13112}}].

\bibitem{Braun:2020mhk}
{\scshape QCD} collaboration, \emph{{Chiral and effective $U(1)_{\rm A}$
  symmetry restoration in QCD}},
  \href{https://arxiv.org/abs/2012.06231}{{\ttfamily 2012.06231}}.

\bibitem{Karsch:2001nf}
F.~Karsch, E.~Laermann and C.~Schmidt, \emph{{The Chiral critical point in
  three-flavor QCD}},
  \href{https://doi.org/10.1016/S0370-2693(01)01114-5}{\emph{Phys. Lett.}
  {\bfseries B520} (2001) 41}
  [\href{https://arxiv.org/abs/hep-lat/0107020}{{\ttfamily hep-lat/0107020}}].

\bibitem{deForcrand:2003vyj}
P.~de~Forcrand and O.~Philipsen, \emph{{The QCD phase diagram for three
  degenerate flavors and small baryon density}},
  \href{https://doi.org/10.1016/j.nuclphysb.2003.09.005}{\emph{Nucl. Phys.}
  {\bfseries B673} (2003) 170}
  [\href{https://arxiv.org/abs/hep-lat/0307020}{{\ttfamily hep-lat/0307020}}].

\bibitem{Iwasaki:1996zt}
Y.~Iwasaki, K.~Kanaya, S.~Kaya, S.~Sakai and T.~Yoshie, \emph{{Finite
  temperature transitions in lattice QCD with Wilson quarks: Chiral transitions
  and the influence of the strange quark}},
  \href{https://doi.org/10.1103/PhysRevD.54.7010}{\emph{Phys. Rev. D}
  {\bfseries 54} (1996) 7010}
  [\href{https://arxiv.org/abs/hep-lat/9605030}{{\ttfamily hep-lat/9605030}}].

\bibitem{Jin:2014hea}
X.-Y.~Jin, Y.~Kuramashi, Y.~Nakamura, S.~Takeda and A.~Ukawa, \emph{{Critical
  endpoint of the finite temperature phase transition for three flavor QCD}},
  \href{https://doi.org/10.1103/PhysRevD.91.014508}{\emph{Phys. Rev. D}
  {\bfseries 91} (2015) 014508}
  [\href{https://arxiv.org/abs/1411.7461}{{\ttfamily 1411.7461}}].

\bibitem{Bonati:2014kpa}
C.~Bonati, P.~de~Forcrand, M.~D'Elia, O.~Philipsen and F.~Sanfilippo,
  \emph{{Chiral phase transition in two-flavor QCD from an imaginary chemical
  potential}}, \href{https://doi.org/10.1103/PhysRevD.90.074030}{\emph{Phys.
  Rev.} {\bfseries D90} (2014) 074030}
  [\href{https://arxiv.org/abs/1408.5086}{{\ttfamily 1408.5086}}].

\bibitem{Cuteri:2017gci}
F.~Cuteri, O.~Philipsen and A.~Sciarra, \emph{{QCD chiral phase transition from
  noninteger numbers of flavors}},
  \href{https://doi.org/10.1103/PhysRevD.97.114511}{\emph{Phys. Rev. D}
  {\bfseries 97} (2018) 114511}
  [\href{https://arxiv.org/abs/1711.05658}{{\ttfamily 1711.05658}}].

\bibitem{Philipsen:2016hkv}
O.~Philipsen and C.~Pinke, \emph{{The $N_f=2$ QCD chiral phase transition with
  Wilson fermions at zero and imaginary chemical potential}},
  \href{https://doi.org/10.1103/PhysRevD.93.114507}{\emph{Phys. Rev.}
  {\bfseries D93} (2016) 114507}
  [\href{https://arxiv.org/abs/1602.06129}{{\ttfamily 1602.06129}}].

\bibitem{Bazavov:2017xul}
A.~Bazavov, H.T.~Ding, P.~Hegde, F.~Karsch, E.~Laermann, S.~Mukherjee et~al.,
  \emph{{Chiral phase structure of three flavor QCD at vanishing baryon number
  density}}, \href{https://doi.org/10.1103/PhysRevD.95.074505}{\emph{Phys. Rev.
  D} {\bfseries 95} (2017) 074505}
  [\href{https://arxiv.org/abs/1701.03548}{{\ttfamily 1701.03548}}].

\bibitem{Jin:2017jjp}
X.-Y.~Jin, Y.~Kuramashi, Y.~Nakamura, S.~Takeda and A.~Ukawa, \emph{{Critical
  point phase transition for finite temperature 3-flavor QCD with
  non-perturbatively O($a$) improved Wilson fermions at $N_{\rm t}=10$}},
  \href{https://doi.org/10.1103/PhysRevD.96.034523}{\emph{Phys. Rev.}
  {\bfseries D96} (2017) 034523}
  [\href{https://arxiv.org/abs/1706.01178}{{\ttfamily 1706.01178}}].

\bibitem{Kuramashi:2020meg}
Y.~Kuramashi, Y.~Nakamura, H.~Ohno and S.~Takeda, \emph{{Nature of the phase
  transition for finite temperature $N_{\rm f}=3$ QCD with nonperturbatively
  O($a$) improved Wilson fermions at $N_{\rm t}=12$}},
  \href{https://doi.org/10.1103/PhysRevD.101.054509}{\emph{Phys. Rev. D}
  {\bfseries 101} (2020) 054509}
  [\href{https://arxiv.org/abs/2001.04398}{{\ttfamily 2001.04398}}].

\bibitem{deForcrand:2017cgb}
P.~de~Forcrand and M.~D'Elia, \emph{{Continuum limit and universality of the
  Columbia plot}}, {\emph{PoS} {\bfseries LATTICE2016} (2017) 081}
  [\href{https://arxiv.org/abs/1702.00330}{{\ttfamily 1702.00330}}].

\bibitem{Ohno:2018gcx}
H.~Ohno, Y.~Kuramashi, Y.~Nakamura and S.~Takeda, \emph{{Continuum
  extrapolation of the critical endpoint in 4-flavor QCD with Wilson-Clover
  fermions}}, \href{https://doi.org/10.22323/1.334.0174}{\emph{PoS} {\bfseries
  LATTICE2018} (2018) 174} [\href{https://arxiv.org/abs/1812.01318}{{\ttfamily
  1812.01318}}].

\bibitem{Ding:2019prx}
H.T.~Ding et~al., \emph{{Chiral Phase Transition Temperature in ( 2+1 )-Flavor
  QCD}}, \href{https://doi.org/10.1103/PhysRevLett.123.062002}{\emph{Phys. Rev.
  Lett.} {\bfseries 123} (2019) 062002}
  [\href{https://arxiv.org/abs/1903.04801}{{\ttfamily 1903.04801}}].

\bibitem{Kotov:2021rah}
A.Y.~Kotov, M.P.~Lombardo and A.~Trunin, \emph{{QCD transition at the physical
  point, and its scaling window from twisted mass Wilson fermions}},
  \href{https://arxiv.org/abs/2105.09842}{{\ttfamily 2105.09842}}.

\bibitem{lawrie}
I.~Lawrie and S.~Sarbach, \emph{{Theory of tricritical points}},  in
  \emph{Phase transitions and critical phenomena}, C.~Domb and J.~Lebowitz,
  eds., vol.~9, p.~1 (1984).

\bibitem{deForcrand:2006pv}
P.~de~Forcrand and O.~Philipsen, \emph{{The Chiral critical line of N(f) = 2+1
  QCD at zero and non-zero baryon density}},
  \href{https://doi.org/10.1088/1126-6708/2007/01/077}{\emph{JHEP} {\bfseries
  0701} (2007) 077} [\href{https://arxiv.org/abs/hep-lat/0607017}{{\ttfamily
  hep-lat/0607017}}].

\bibitem{DeGrand:2015zxa}
T.~DeGrand, \emph{{Lattice tests of beyond Standard Model dynamics}},
  \href{https://doi.org/10.1103/RevModPhys.88.015001}{\emph{Rev. Mod. Phys.}
  {\bfseries 88} (2016) 015001}
  [\href{https://arxiv.org/abs/1510.05018}{{\ttfamily 1510.05018}}].

\bibitem{Nogradi:2016qek}
D.~Nogradi and A.~Patella, \emph{{Strong dynamics, composite Higgs and the
  conformal window}},
  \href{https://doi.org/10.1142/S0217751X1643003X}{\emph{Int. J. Mod. Phys. A}
  {\bfseries 31} (2016) 1643003}
  [\href{https://arxiv.org/abs/1607.07638}{{\ttfamily 1607.07638}}].

\bibitem{Svetitsky:2017xqk}
B.~Svetitsky, \emph{{Looking behind the Standard Model with lattice gauge
  theory}}, \href{https://doi.org/10.1051/epjconf/201817501017}{\emph{EPJ Web
  Conf.} {\bfseries 175} (2018) 01017}
  [\href{https://arxiv.org/abs/1708.04840}{{\ttfamily 1708.04840}}].

\bibitem{Braun:2006jd}
J.~Braun and H.~Gies, \emph{{Chiral phase boundary of QCD at finite
  temperature}},
  \href{https://doi.org/10.1088/1126-6708/2006/06/024}{\emph{JHEP} {\bfseries
  06} (2006) 024} [\href{https://arxiv.org/abs/hep-ph/0602226}{{\ttfamily
  hep-ph/0602226}}].

\bibitem{Braun:2009ns}
J.~Braun and H.~Gies, \emph{{Scaling laws near the conformal window of
  many-flavor QCD}}, \href{https://doi.org/10.1007/JHEP05(2010)060}{\emph{JHEP}
  {\bfseries 05} (2010) 060} [\href{https://arxiv.org/abs/0912.4168}{{\ttfamily
  0912.4168}}].

\bibitem{Miura:2012zqa}
K.~Miura and M.P.~Lombardo, \emph{{Lattice Monte-Carlo study of pre-conformal
  dynamics in strongly flavoured QCD in the light of the chiral phase
  transition at finite temperature}},
  \href{https://doi.org/10.1016/j.nuclphysb.2013.02.008}{\emph{Nucl. Phys. B}
  {\bfseries 871} (2013) 52} [\href{https://arxiv.org/abs/1212.0955}{{\ttfamily
  1212.0955}}].

\bibitem{pinke_cl2qcd_2018}
C.~Pinke, M.~Bach, A.~Sciarra, F.~Cuteri, L.~Zeidlewicz, C.~Sch\"afer et~al.,
  \emph{{\texttt{CL\textsuperscript{2}QCD~--~v1.0}}},  Sept., 2018.
\newblock 10.5281/zenodo.5121895.

\bibitem{sciarra_cl2qcd_2021}
A.~Sciarra, C.~Pinke, M.~Bach, F.~Cuteri, L.~Zeidlewicz, C.~Sch\"afer et~al.,
  \emph{{\texttt{CL\textsuperscript{2}QCD~--~v1.1}}},  Feb., 2021.
\newblock 10.5281/zenodo.5121917.

\bibitem{sciarra_bahamas_2021}
A.~Sciarra, \emph{{\texttt{BaHaMAS}}},  Feb., 2021.
\newblock 10.5281/zenodo.4577425.

\bibitem{Clark2007}
M.A.~Clark and A.D.~Kennedy, \emph{Accelerating dynamical-fermion computations
  using the rational hybrid monte carlo algorithm with multiple pseudofermion
  fields}, \href{https://doi.org/10.1103/physrevlett.98.051601}{\emph{Physical
  Review Letters} {\bfseries 98} (2007) }.

\bibitem{Binder:1981sa}
K.~Binder, \emph{{Finite size scaling analysis of Ising model block
  distribution functions}},
  \href{https://doi.org/10.1007/BF01293604}{\emph{Z.Phys.} {\bfseries B43}
  (1981) 119}.

\bibitem{Pelissetto:2000ek}
A.~Pelissetto and E.~Vicari, \emph{{Critical phenomena and renormalization
  group theory}},
  \href{https://doi.org/10.1016/S0370-1573(02)00219-3}{\emph{Phys. Rept.}
  {\bfseries 368} (2002) 549}
  [\href{https://arxiv.org/abs/cond-mat/0012164}{{\ttfamily
  cond-mat/0012164}}].

\bibitem{Cuteri:2018wci}
F.~Cuteri, O.~Philipsen and A.~Sciarra, \emph{{Progress on the nature of the
  QCD thermal transition as a function of quark flavors and masses}},
  \href{https://doi.org/10.22323/1.334.0170}{\emph{PoS} {\bfseries LATTICE2018}
  (2018) 170} [\href{https://arxiv.org/abs/1811.03840}{{\ttfamily
  1811.03840}}].

\bibitem{FSReweighting}
A.M.~Ferrenberg and R.H.~Swendsen, \emph{Optimized monte carlo data analysis},
  \href{https://doi.org/10.1103/PhysRevLett.63.1195}{\emph{Phys. Rev. Lett.}
  {\bfseries 63} (1989) 1195}.

\bibitem{deForcrand:2007rq}
P.~de~Forcrand, S.~Kim and O.~Philipsen, \emph{{A QCD chiral critical point at
  small chemical potential: Is it there or not?}}, {\emph{PoS} {\bfseries
  LAT2007} (2007) 178} [\href{https://arxiv.org/abs/0711.0262}{{\ttfamily
  0711.0262}}].

\bibitem{Sharpe:2006re}
S.R.~Sharpe, \emph{{Rooted staggered fermions: Good, bad or ugly?}},
  \href{https://doi.org/10.22323/1.032.0022}{\emph{PoS} {\bfseries LAT2006}
  (2006) 022} [\href{https://arxiv.org/abs/hep-lat/0610094}{{\ttfamily
  hep-lat/0610094}}].

\bibitem{Kronfeld:2007ek}
A.S.~Kronfeld, \emph{{Lattice Gauge Theory with Staggered Fermions: How, Where,
  and Why (Not)}}, \href{https://doi.org/10.22323/1.042.0016}{\emph{PoS}
  {\bfseries LATTICE2007} (2007) 016}
  [\href{https://arxiv.org/abs/0711.0699}{{\ttfamily 0711.0699}}].

\bibitem{Cheng:2011ic}
A.~Cheng, A.~Hasenfratz and D.~Schaich, \emph{{Novel phase in SU(3) lattice
  gauge theory with 12 light fermions}},
  \href{https://doi.org/10.1103/PhysRevD.85.094509}{\emph{Phys. Rev. D}
  {\bfseries 85} (2012) 094509}
  [\href{https://arxiv.org/abs/1111.2317}{{\ttfamily 1111.2317}}].

\bibitem{Kotov:2021mgp}
A.Y.~Kotov, D.~Nogradi, K.K.~Szabo and L.~Szikszai, \emph{{More on the flavor
  dependence of $m_\varrho / f_\pi$}},
  \href{https://arxiv.org/abs/2107.05996}{{\ttfamily 2107.05996}}.

\bibitem{Varnhorst:2015lea}
L.~Varnhorst, \emph{{The $N_f$=3 critical endpoint with smeared staggered
  quarks}}, \href{https://doi.org/10.22323/1.214.0193}{\emph{PoS} {\bfseries
  LATTICE2014} (2015) 193}.

\bibitem{Bochicchio:1985xa}
M.~Bochicchio, L.~Maiani, G.~Martinelli, G.C.~Rossi and M.~Testa, \emph{{Chiral
  Symmetry on the Lattice with Wilson Fermions}},
  \href{https://doi.org/10.1016/0550-3213(85)90290-1}{\emph{Nucl. Phys. B}
  {\bfseries 262} (1985) 331}.

\bibitem{Aoki:1983qi}
S.~Aoki, \emph{{New Phase Structure for Lattice QCD with Wilson Fermions}},
  \href{https://doi.org/10.1103/PhysRevD.30.2653}{\emph{Phys. Rev. D}
  {\bfseries 30} (1984) 2653}.

\bibitem{Aoki:1986xr}
S.~Aoki, \emph{{A Solution to the U(1) Problem on a Lattice}},
  \href{https://doi.org/10.1103/PhysRevLett.57.3136}{\emph{Phys. Rev. Lett.}
  {\bfseries 57} (1986) 3136}.

\bibitem{Bitar:1996kc}
K.M.~Bitar, \emph{{Absence of parity flavor breaking phase in QCD with two
  flavors of Wilson fermions for Beta =\ensuremath{>} 5.0}},
  \href{https://doi.org/10.1103/PhysRevD.56.2736}{\emph{Phys. Rev. D}
  {\bfseries 56} (1997) 2736}
  [\href{https://arxiv.org/abs/hep-lat/9602027}{{\ttfamily hep-lat/9602027}}].

\bibitem{Ilgenfritz:2003gw}
E.-M.~Ilgenfritz, W.~Kerler, M.~Muller-Preussker, A.~Sternbeck and H.~Stuben,
  \emph{{A Numerical reinvestigation of the Aoki phase with N(f) = 2 Wilson
  fermions at zero temperature}},
  \href{https://doi.org/10.1103/PhysRevD.69.074511}{\emph{Phys. Rev. D}
  {\bfseries 69} (2004) 074511}
  [\href{https://arxiv.org/abs/hep-lat/0309057}{{\ttfamily hep-lat/0309057}}].

\bibitem{Sharpe:1998xm}
S.R.~Sharpe and R.L.~Singleton, Jr, \emph{{Spontaneous flavor and parity
  breaking with Wilson fermions}},
  \href{https://doi.org/10.1103/PhysRevD.58.074501}{\emph{Phys. Rev. D}
  {\bfseries 58} (1998) 074501}
  [\href{https://arxiv.org/abs/hep-lat/9804028}{{\ttfamily hep-lat/9804028}}].

\bibitem{Munster:2004am}
G.~Munster, \emph{{On the phase structure of twisted mass lattice QCD}},
  \href{https://doi.org/10.1088/1126-6708/2004/09/035}{\emph{JHEP} {\bfseries
  09} (2004) 035} [\href{https://arxiv.org/abs/hep-lat/0407006}{{\ttfamily
  hep-lat/0407006}}].

\bibitem{Farchioni:2004us}
F.~Farchioni, R.~Frezzotti, K.~Jansen, I.~Montvay, G.C.~Rossi, E.~Scholz
  et~al., \emph{{Twisted mass quarks and the phase structure of lattice QCD}},
  \href{https://doi.org/10.1140/epjc/s2004-02078-9}{\emph{Eur. Phys. J. C}
  {\bfseries 39} (2005) 421}
  [\href{https://arxiv.org/abs/hep-lat/0406039}{{\ttfamily hep-lat/0406039}}].

\bibitem{Blum:1994eh}
T.~Blum, T.A.~DeGrand, C.E.~Detar, S.A.~Gottlieb, A.~Hasenfratz, L.~Karkkainen
  et~al., \emph{{QCD thermodynamics with Wilson quarks at large kappa}},
  \href{https://doi.org/10.1103/PhysRevD.50.3377}{\emph{Phys. Rev. D}
  {\bfseries 50} (1994) 3377}
  [\href{https://arxiv.org/abs/hep-lat/9404006}{{\ttfamily hep-lat/9404006}}].

\bibitem{Ilgenfritz:2009ns}
E.M.~Ilgenfritz, K.~Jansen, M.P.~Lombardo, M.~Muller-Preussker, M.~Petschlies,
  O.~Philipsen et~al., \emph{{Phase structure of thermal lattice QCD with N(f)
  = 2 twisted mass Wilson fermions}},
  \href{https://doi.org/10.1103/PhysRevD.80.094502}{\emph{Phys. Rev. D}
  {\bfseries 80} (2009) 094502}
  [\href{https://arxiv.org/abs/0905.3112}{{\ttfamily 0905.3112}}].

\bibitem{Aoki:2004iq}
{\scshape JLQCD} collaboration, \emph{{Bulk first-order phase transition in
  three-flavor lattice QCD with O(a)-improved Wilson fermion action at zero
  temperature}}, \href{https://doi.org/10.1103/PhysRevD.72.054510}{\emph{Phys.
  Rev. D} {\bfseries 72} (2005) 054510}
  [\href{https://arxiv.org/abs/hep-lat/0409016}{{\ttfamily hep-lat/0409016}}].

\bibitem{Iwasaki:2011np}
Y.~Iwasaki, \emph{{Renormalization Group Analysis of Lattice Theories and
  Improved Lattice Action. II. Four-dimensional non-Abelian SU(N) gauge
  model}},  \href{https://arxiv.org/abs/1111.7054}{{\ttfamily 1111.7054}}.

\bibitem{Aoki:2005et}
{\scshape CP-PACS, JLQCD} collaboration, \emph{{Nonperturbative O(a)
  improvement of the Wilson quark action with the RG-improved gauge action
  using the Schr\"odinger functional method}},
  \href{https://doi.org/10.1103/PhysRevD.73.034501}{\emph{Phys. Rev. D}
  {\bfseries 73} (2006) 034501}
  [\href{https://arxiv.org/abs/hep-lat/0508031}{{\ttfamily hep-lat/0508031}}].

\bibitem{AliKhan:2000wou}
{\scshape CP-PACS} collaboration, \emph{{Phase structure and critical
  temperature of two flavor QCD with renormalization group improved gauge
  action and clover improved Wilson quark action}},
  \href{https://doi.org/10.1103/PhysRevD.63.034502}{\emph{Phys. Rev. D}
  {\bfseries 63} (2000) 034502}
  [\href{https://arxiv.org/abs/hep-lat/0008011}{{\ttfamily hep-lat/0008011}}].

\bibitem{Brandt:2016daq}
B.B.~Brandt, A.~Francis, H.B.~Meyer, O.~Philipsen, D.~Robaina and H.~Wittig,
  \emph{{On the strength of the $U_A(1)$ anomaly at the chiral phase transition
  in $N_f=2$ QCD}}, \href{https://doi.org/10.1007/JHEP12(2016)158}{\emph{JHEP}
  {\bfseries 12} (2016) 158}
  [\href{https://arxiv.org/abs/1608.06882}{{\ttfamily 1608.06882}}].

\bibitem{Brandt:2019ksy}
B.B.~Brandt, O.~Philipsen, M.~C\`e, A.~Francis, T.~Harris, H.B.~Meyer et~al.,
  \emph{{Testing the strength of the $\text{U}_A(1)$ anomaly at the chiral
  phase transition in two-flavour QCD}},
  \href{https://doi.org/10.22323/1.317.0055}{\emph{PoS} {\bfseries CD2018}
  (2019) 055} [\href{https://arxiv.org/abs/1904.02384}{{\ttfamily
  1904.02384}}].

\bibitem{Friedemann_2017}
S.~Friedemann, W.J.~Duncan, M.~Hirschberger, T.W.~Bauer, R.~Küchler,
  A.~Neubauer et~al., \emph{{Quantum tricritical points in NbFe$_2$}},
  \href{https://doi.org/10.1038/nphys4242}{\emph{Nature Physics} {\bfseries 14}
  (2017) 62}.

\bibitem{Tetradis:1995br}
N.~Tetradis and D.F.~Litim, \emph{{Analytical solutions of exact
  renormalization group equations}},
  \href{https://doi.org/10.1016/0550-3213(95)00642-7}{\emph{Nucl. Phys. B}
  {\bfseries 464} (1996) 492}
  [\href{https://arxiv.org/abs/hep-th/9512073}{{\ttfamily hep-th/9512073}}].

\bibitem{DelDebbio:2011zz}
L.~Del~Debbio and L.~Keegan, \emph{{RG flows in 3D scalar field theory}},
  \href{https://doi.org/10.22323/1.139.0061}{\emph{PoS} {\bfseries LATTICE2011}
  (2011) 061}.

\end{thebibliography}\endgroup

\end{document}